\documentclass[11pt]{article}
\usepackage{amsmath,sint,epsfig,cite,graphics}
\usepackage{macros_static}
\usepackage{textcomp}
\usepackage{wrapfig}

\begin{document}

\begin{titlepage}

\begin{flushright}
\vskip 0.7cm
DESY 06-165 \\
HU-EP-06/19\\
RM3-TH/06-15\\
SFB/CPP-06-34\\
\end{flushright}

\vskip 0.95cm
\begin{center}
{\Large\bf 
Heavy Quark Effective Theory computation of the mass of the bottom quark.
\\[0.5ex] 
}
\end{center}
\vskip 0.35cm
\vbox{
\centerline{
\epsfxsize=2.8 true cm
\epsfbox{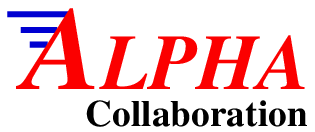}}
}
\vskip 0.1cm
\begin{center}
{
Michele Della Morte$^{\scriptscriptstyle a}$,
Nicolas Garron$^{\scriptscriptstyle b}$,
Mauro Papinutto$^{\scriptscriptstyle c}$ and  
Rainer Sommer$^{\scriptscriptstyle b}$
}
\vskip 0.5cm
{
$^{\scriptstyle a}$
Institut f\"ur Physik, Humboldt Universit\"at, \\
Newtonstr. 15, 12489 Berlin, Germany 
\vskip 2.0ex
$^{\scriptstyle b}$
DESY,
Platanenallee 6, 15738 Zeuthen,  Germany
\vskip 2.0ex
$^{\scriptstyle c}$
INFN Sezione di Roma Tre, Via della Vasca Navale 84, 
I-00146, Rome, Italy
\vskip 2.0ex
}
\vskip 0.775cm
{\bf Abstract}
\vskip 0.1ex
\end{center}
We present a fully non-perturbative 
computation of the 
mass of the b-quark in the 
quenched approximation. Our strategy starts from the 
matching of HQET to QCD in a finite volume and finally
relates the quark mass to
the spin averaged mass of the $\rm B_\mrm{s}$ meson in HQET.
All steps include the terms of order  $\Lambda^2/\mbeauty$. 
Expanding on \cite{hqet:pap1}, we discuss the computation 
and renormalization of correlation functions at order $\minv$. 
With the strange quark mass fixed from  
the Kaon mass and the QCD scale set through $r_0=0.5\,\fm$, 
we obtain a renormalization group invariant
mass $\Mbeauty =  6.758(86)\, \GeV$ or 
$ \mbar_\beauty(\mbar_\beauty)= 4.347(48) \GeV$
in the $\msbar$ scheme. The uncertainty in the computed 
$\Lambda^2/\mbeauty$ terms contributes little to the 
total error and $\Lambda^3/\mbeauty^2$ terms are negligible.
The strategy is promising for full QCD as well as for other
B-physics observables.

\vskip 0.35cm
\vfill

\begin{center}
September 2006
\end{center}

\eject
\vfill
\eject

\end{titlepage}

\section{Introduction \label{s:intro}}

The mass of the b-quark, $\Mbeauty$, is a relevant input parameter
for phenomenology analysis based on perturbation theory.
Let us just mention the extraction of $V_\mrm{ub}$ from inclusive 
b-decays \cite{Bigi:inclusive,CKM:2003}. $\Mbeauty$ is a fundamental
parameter of the Standard Model of particle physics. Thus it should 
be determined precisely. 
One may of course turn the very first observation around. 
For instance, applying high order perturbation theory to sufficiently 
well integrated cross sections, the quark mass can be 
determined~\cite{mb:melnikov,mb:beneke,mb:hoang00,mb:steinhkuehn,mb:erler,mb:eidemuller,mb:corcella,mb:bordes,Pineda:2006gx,Boughezal:2006px,Penin:1998zh,Penin:1998kx}.
Still, the achievable precision is limited 
by the intrinsic uncertainty of perturbation theory
and maybe more by experimental difficulties. 
On the other hand, the use of lattice QCD offers a strategy 
to compute the fundamental renormalization group invariant (RGI)
parameters of QCD with very precisely known experimental
input, e.g. $\nf$ (=~number of quarks flavours) meson masses 
as well as the nucleon mass; see 
\cite{reviews:fund} for a basic introduction. 
However, for the b-quark mass, such a computation is more
involved than for the light quarks because the achievable inverse lattice
spacings are below the mass of the quark. Effective theories have to
be employed in a numerical treatment of the bound states. The most 
serious problem that arises is that a power law divergent ($\sim g_0^2/a$) 
additive renormalization of the mass is present due to
the absence of a chiral symmetry in the effective theories
(even in the continuum). Although at the 
lowest order in Heavy Quark Effective Theory (HQET) the subtraction is known to 
order $g_0^6/a$ \cite{stat:eichhill2,mbstat:MaSa,stat:deltamnnlo}, 
an (in the continuum limit)
divergent remainder is unavoidable and the total uncertainty 
is difficult to estimate as long as the renormalization is carried
out perturbatively.

In \cite{hqet:pap1} a general strategy was described which 
allows HQET at zero velocity to be implemented non-perturbatively 
on the lattice, including all renormalizations.

%
\newcommand{\ftext}[1]{\fbox{ {#1} }}

\vspace{2.5cm}\hspace{1.5cm}
\begin{picture}(8,20)(0,0)
\small
  \unitlength 0.4cm
  \put(2,6){\ftext{experiment}}            \put(16.5,6){\ftext{Lattice with 
$a\mq\ll 1$}} 
  \put(2,4){ $\mB=5.4\,\GeV$}    \put(16,4){ $\Phi_1(L_1,\Mbeauty),\Phi_2(L_1,\Mbeauty)$} 
  \linethickness{0.3mm}\put(5.3,3.5){\vector(0,-1){1.5}}
  \linethickness{0.3mm}\put(21,3.5){\vector(0,-1){1.5}}
  \linethickness{0.3mm}
  \put(0,0.5){ $\Phi_1^\mrm{HQET}(L_2),\Phi_2^\mrm{HQET}(L_2)$}
  \put(16,0.5){ $\Phi_1^\mrm{HQET}(L_1),\Phi_2^\mrm{HQET}(L_1)$}
  \put(14.7,0.7){\vector(-1,0){5.0}}
  \put(10.5,1.1){$\sigmam(u_1)$}
  \put(9.5,-0.4){$\sigmakin_1(u_1),\sigmakin_2(u_1)$}
  \put(10,3.0){\small $L_2 = 2 L_1$}
\end{picture}
\vspace{0.3cm}


The basic idea is illustrated in the above diagram. 
It is founded on the knowledge 
of the relation between the RGI mass and the bare mass in QCD 
\cite{mbar:pap1,mbar:nf2}.
In a finite volume of extent 
$L_1\approx 0.4\,\fm$, one chooses lattice spacings 
$a$ sufficiently smaller than $\minv$, such that the b-quark 
propagates correctly up to controllable discretization errors
of order $a^2$. Finite volume observables $\Phi_i(L_1,\Mbeauty)$
may then be computed as a function of the RGI mass $\Mbeauty$ including
an extrapolation to the continuum limit. The resulting values
are equated to
their representation in HQET -- a step called matching, indicated
by the r.h.s. of the diagram. 
Choosing now $L_1/a=\rmO(10)$, with the same physical value of $L_1$,
one uses the knowledge of
$\Phi_i(L_1,\Mbeauty)$ to determine 
the bare parameters in the effective theory 
for $a$-values of about $0.025\,\fm$ to $0.05\,\fm$. 
At these lattice spacings one then computes
the same observables in a larger volume $L_2=2L_1$. Again
these observables can be extrapolated to the continuum limit.
Next, the knowledge of $\Phi_i(L_2,\Mbeauty)$ and the choice 
$L_2/a=\rmO(10)$
yields the bare parameters of the effective theory
for $a$ around $0.05\,\fm$ to $0.1\,\fm$. 
One then has full control over the
effective theory at lattice spacings where large volume
observables, such as the B-meson mass, can be computed.
Perturbation theory is completely avoided with the power divergent 
subtractions being taken care of non-perturbatively.

We return to the specific application of computing $\Mbeauty$.
The whole chain allows to express 
$\mB$ in terms of $\Phi_i(L_1,\Mbeauty)$
and thus as a function of $\Mbeauty$. 
This function naturally splits into various pieces which may be
computed individually as they separately have a continuum limit. 
In particular, the step scaling
functions $\sigma$ relate $\Phi_i(L_1)$ to $\Phi_i(L_2)$. 
As we will see below,
at first order in $\minv$,
two matching observables $\Phi_1,\Phi_2$
are sufficient if we consider the spin averaged B-meson mass.

The strategy requires all considered observables to be accurately  
described by the $\minv$ expansion. 
Naive counting estimates the accuracy of the quark mass as
$\Lambda \times {\Lambda^2 \over \mbeauty^2}$ and 
$\Lambda \times {1 \over L_1^2 \mbeauty^2}$. For 
a typical QCD scale $\Lambda\approx 400\,\MeV$
both these terms yield the same, very small, estimate. In \cite{hqet:pap3} the 
${\minv}$ expansion was tested for an even smaller $L=L_0=L_1/2$
and found to be well behaved, as it is also the case in perturbation theory 
\cite{zastat:pap2}.
Here we will have additional cross
checks by choosing different quantities $\Phi_i$ in the matching step.

In \sect{s:hqet} we will 
go through the definition of the effective theory
in order to fix some notations and 
give rules how the $\minv$ expansion is
implemented in practice.
We also discuss
correlation functions in the \SF \cite{SF:LNWW,SF:stefan1},
which defines our finite volume geometry.
These correlation functions are then used in \sect{s:phi} 
to form suitable dimensionless observables  $\Phi_i$,
followed by a section which lists the step scaling
functions. \sect{s:mb} discusses the final formula for the
RGI b-quark mass $\Mbeauty$. 
Numerical results for all quantities in the quenched approximation
are discussed in \sect{s:res}. This includes also results from
an alternative strategy as a check on the smallness of the $\minv^2$ 
terms.


\section{Heavy quark effective theory on the lattice \label{s:hqet}}

We start from the Eichten Hill static quark Lagrangian~\cite{stat:eichhill1},
using the notation of \cite{stat:actpaper}, but setting the
mass counter term $\dmstat$ 
to zero. Its effect is taken into account in the overall
energy shift $\mhbare$ between the effective theory and 
QCD. Thus $\mhbare$ is regularization dependent 
with a $\sim g_0^2/a$ divergence. 
For the sake of a light notation, we also drop the 
superscript W \cite{stat:actpaper} for the different lattice discretizations of the 
static Lagrangian, but in the numerical computations these 
different versions will be used
and referred to exactly as in that reference. We remind the reader
that they differ only by the choice of the covariant derivative $D_0$.

The terms of first order in $\minv$ are introduced exactly as
in \cite{hqet:pap1}, but we use a slightly different notation which is
convenient when one does not go beyond that order.

\subsection{Formulation}

The lowest order (static) Lagrangian, 
\be
\label{e:statact}
\lag{stat}(x) = 
\heavyb(x) \,D_0\, \heavy(x) \;,
\ee
is written in terms of the 
backward covariant derivative $D_0$ as in \cite{stat:actpaper}
and
the 4-component  heavy quark field subject to the constraints 
$P_+ \heavy = \heavy \,, \; \heavyb P_+=\heavyb$ with
$P_+={(1+\gamma_0)/{2}}$. 
At the first order we write the HQET Lagrangian 
\bes
\lag{HQET}(x) &=&  \lag{stat}(x) + \lnu{1}(x) \,,\\
  \lnu{1}(x) &=& - \omegaspin\Ospin(x) 
        - \omegakin\Okin(x) \,,\\
  \Ospin(x) &=& \heavyb(x){\boldsymbol\sigma}\!\cdot\!{\bf B}\heavy(x)\,\quad 
    \Okin  = \heavyb(x){\bf D}^2\heavy(x) \,,
\ees
such that the {\em classical} values for the coefficients
are $\omegakin=\omegaspin=1/(2\mbeauty)$. We use the discretized version
$
 {\boldsymbol\sigma}\!\cdot\!{\bf B} = \sum_{k,j}\sigma_{kj}\widehat{F}_{kj}/(2i)\,,
$
with $\sigma_{kj}$ and the lattice field tensor $\widehat{F}$ defined
in \cite{zastat:pap1}. The kinetic term ${\bf D}^2$ is represented 
by the nearest neighbor covariant 3-d Laplacian. 
The effective theory is expected to be renormalizable at each 
(fixed) order in $\minv$ if (and only if) path integral expectation values 
are defined by expanding the path integral weight as~\cite{hqet:pap1}
\bes
  \exp(-a^4\sum_x[\lag{HQET}(x)+\lag{light}(x)]) &=&  
  \exp(-a^4\sum_x[\lag{stat}(x)+\lag{light}(x)])\\
        &\times& \Big (
        1 + a^4\sum_x [\omegaspin\Ospin(x) + \omegakin\Okin(x)]\Big )\,. 
        \nonumber
\ees
For correlation functions of some multilocal 
fields $\op{}$ this means
\bes
  \label{e:expval}
  \langle \op{} \rangle &=& 
         \langle  \op{}  \rangle_\mrm{stat} 
        + \omegakin  a^4\sum_x \langle  \op{} \Okin(x) \rangle_\mrm{stat}
        + \omegaspin a^4\sum_x \langle \op{} \Ospin(x) \rangle_\mrm{stat} \\
  &\equiv&  \langle  \op{}  \rangle_\mrm{stat} 
        + \omegakin\langle  \op{}  \rangle_\mrm{kin} 
        + \omegaspin\langle  \op{}  \rangle_\mrm{spin} \,,
        \label{e:exp}
\ees
where $\langle  \op{}  \rangle_\mrm{stat}$ denotes the static
expectation value with Lagrangian $\lag{stat}(x) + \lag{light}(x)$. 
All terms composed of just the relativistic quarks and the gauge fields 
are summarized 
in $\lag{light}(x)$. Note that as one performs the Wick contractions of
the heavy quark field,
the $\minv$ terms $\Okin(x),\Ospin(x)$ leave behind insertions in the 
static heavy quark propagators. From the point of view of renormalization
all terms in \eq{e:expval} are correlation functions in the static effective theory, 
which is power counting renormalizable. 

The above form assumes that $\op{}$ contains all 
$\minv$ terms needed to represent the local fields
in the effective theory. A relevant
example is the time component of the 
heavy light axial current. 
In the effective theory it is represented as
\bes
  \label{e:ahqet}
 \Ahqet(x)&=& \zahqet\,[\Astat(x)+ \cahqet\delta\Astat(x)]\,, \\
 \Astat(x)&=&\lightb(x)\gamma_0\gamma_5\heavy(x) \,,\quad \\
 \delta\Astat(x) &=& \lightb(x){1\over2}
            (\lnab{i}\!+\!\lnabstar{i})\gamma_i\gamma_5\heavy(x)\,.
\ees
Later we will also use the space components of the 
vector current represented by
\bes
  \label{e:vhqet}
 \Vkhqet(x)&=& \zvhqet\,[\Vkstat(x) + \cvhqet\delta\Vkstat(x)]\,,\\
 \Vkstat(x)&=&\lightb(x)\gamma_k\heavy(x)\;,
        \quad \\
 \delta\Vkstat(x) &=& - \lightb(x){1\over2}
        (\lnab{i}\!+\!\lnabstar{i})\gamma_i\gamma_k\heavy(x) \,.
\ees
We have chosen $\Vkstat,\delta\Vkstat$ such that they are exactly related to 
$\Astat,\delta\Astat$ by a spin rotation. 

The coefficients $\omegakin,\omegaspin,\zahqet,\cahqet,\zvhqet,\cvhqet$ 
are functions of the bare coupling $g_0$ {\em and}
of the heavy quark mass in lattice units. They represent bare parameters
of the effective theory, which are to be fixed by matching to QCD. 
Just like $\omegakin,\omegaspin$, the coefficients $\cahqet,\cvhqet$
are of order $\minv$, while we may write 
\bes
  \zahqet = \zastat + \za^{(1)}\,,\;\mbox{with}\;
  \za^{(1)} = \rmO(\minv)\,,
\ees
and similarly for $\zvhqet$~\footnote{
If $\rmO(a)$ improvement is desired in the static
approximation, there are also $a\,\delta\Astat$,
 $a\,\delta\Vkstat$ corrections 
to the currents.
They are not relevant in the present discussion
but will be taken into account when necessary.}.
Note that in the expansion to first order,
terms such as $\omegakin \cahqet \propto\minv^2$ 
are to be dropped. 

Below we will consider an example and discuss that indeed 
the bare parameters \\
$\omegakin,\omegaspin,\zahqet,\cahqet,\zvhqet,\cvhqet$ and $\mhbare$
 are sufficient to absorb all divergences 
in the effective theory at this order in $\minv$.

\subsection{$\minv$ expansion in a geometry without boundaries}

In order to illustrate further how the expansion works, 
we consider a two-point function of a composite field
in a space-time without boundaries, i.e. with periodic
boundary conditions or in infinite volume. We choose the example
\bes \label{e_caa}
  \caa(x_0) = \za^2 a^3\sum_{\vecx} \Big\langle A_0(x)  (A_0)^{\dagger}(0)
              \Big\rangle
\ees
with the heavy-light axial current in QCD,
$A_\mu=\lightb\gamma_\mu\gamma_5\psi_\beauty$, and $\za$ ensuring
the natural normalization of the current consistent with
current algebra~\cite{boch,impr:pap4}. The
$\minv$ expansion 
reads
\bes
   \label{e:caahqet0}
   \caa(x_0) &=& \rme^{-\mhbare x_0} (\zahqet)^2  a^3\sum_{\vecx}\Big [  
          \langle \Astat(x) (\Astat(0))^\dagger  \rangle_\mrm{stat} \\
        && \;+\,  \omegakin\,\langle  \Astat(x) (\Astat(0))^\dagger  \rangle_\mrm{kin} 
        \,+\,  \omegaspin\langle  \Astat(x) (\Astat(0))^\dagger  \rangle_\mrm{spin} 
        \nonumber\\[0.5ex]
        &&  \;+\, \cahqet\, \langle \Astat(x) (\delta\Astat(0))^\dagger  \rangle_\mrm{stat}
           \,+\, \cahqet\, \langle \delta\Astat(x) (\Astat(0))^\dagger  \rangle_\mrm{stat} \Big] \nonumber \\
        &\equiv& \rme^{-\mhbare x_0} (\zahqet)^2 \,\Big[
        \caa^\mrm{stat}(x_0)+\omegakin\caa^\mrm{kin}(x_0)+\omegaspin\caa^\mrm{spin}(x_0) \nonumber \\ 
        && \qquad \qquad \qquad\qquad\;+ \cahqet[\cdaa^\mrm{stat}(x_0)+\cada^\mrm{stat}(x_0)]
        \Big]
        \label{e:caahqet}
\ees
up to terms of order $\minv^2$. As mentioned in the introduction, 
the mass shift $\mhbare = \rmO(\mbeauty)$ includes an additive mass renormalization.
It is also split up as  
\bes
        \mhbare = \mhbare^\mrm{stat} + \mhbare^{(1)}\,,\;\mbox{with}\;  \mhbare^{(1)}=\rmO(\minv)\,,
\ees
and the expansion $ \rme^{-\mhbare\, x_0} \equiv \rme^{-\mhbare^\mrm{stat}\, x_0}(1-x_0\mhbare^{(1)})$
is understood.

For illustration we check the self consistency of \eq{e:caahqet}. The relevant question concerns
renormalization, namely: are the ``free'' parameters $\mhbare \ldots \cahqet$ sufficient to absorb
all divergences on the r.h.s.? We consider the most difficult term, $\caa^\mrm{kin}(x_0)$.
According to the standard rules of renormalization of composite operators,
it is renormalized as
\bes \label{e:caakinr}
  \Big(\caa^\mrm{kin}\Big)_\mrm{R}(x_0) =  \rme^{-\mhbare^\mrm{stat}\, x_0} 
    \big(\zastat\big)^2  a^7 \sum_{\vecx,\, z}\Big\langle 
    \Astat(x)\, (\Astat(0))^\dagger  \,\big(\Okin\big)_\mrm{R}(z)\Big\rangle_\mrm{stat} + \mbox{C.T.}\,, 
    \nonumber\\[-1ex]
\ees
where C.T. denotes contact terms to be discussed shortly. The renormalized operator $\big(\Okin\big)_\mrm{R}(z)$
involves a subtraction of lower dimensional ones,
\bes \label{e:okinr}
  \big(\Okin\big)_\mrm{R}(z) = Z_{\Okin} \big( \Okin(z) + {c_1\over a}\, \heavyb(z) D_0 \heavy(z) + 
  {c_2\over a^2}\, \heavyb(z)\heavy(z) \big)\,,
\ees
written here in terms of dimensionless $c_i$. Since we are interested in on-shell observables
($x_0>0$ in eq.(2.19)), 
we may use the equation of motion
$D_0 \heavy(z)=0$ to eliminate the second term. The third one, 
${c_2\over a^2} \heavyb(z)\heavy(z)$, is equivalent
to a mass shift and only changes  
$\mhbare^{(1)}$, which is hence quadratically
divergent~\footnote{Using the explicit form of the static propagator, 
eq.~(2.4) of reference \cite{stat:actpaper}, one can check that indeed
$a^3\sum_{\vecx}\,\Big\langle \Astat(x)\, 
  (\Astat(0))^\dagger a^4\sum_z \heavyb(z)\heavy(z) \Big\rangle_\mrm{stat}
  = x_0\caa^\mrm{stat}(x_0)
$. 
}.
 Thus all terms which are needed for the renormalization of $\Okin$ are present 
in \eq{e:caahqet}. It remains to consider the contact terms in \eq{e:caakinr}. 
They originate from singularities
in the operator products
$\Okin(z)\Astat(x)$ as $z\to x$ (and $\Okin(z)\big(\Astat\big)^\dagger(0)$  as $z\to 0$). 
Using the operator product expansion 
they can be represented as linear combinations of $\Astat(x)$ and $\delta\Astat(x)$. Such terms are contained in
\eq{e:caahqet} in the form of $\caa^\mrm{stat},\cdaa^\mrm{stat}$ and $\cada^\mrm{stat}$~\footnote{
$\Astat(x)$ and $\delta\Astat(x)$ are the only operators of dimension 3 and 4 with the correct
quantum numbers.  Higher dimensional operators contribute only terms of order $a$. Note that 
the $\Astat(x)$ term is power divergent $\sim 1/(a\mbeauty)$. This divergence is absorbed
by a power divergent contribution to $\zahqet$ (at order $\minv$).}.

We conclude that all terms which are needed
for the renormalization of $\caa^\mrm{kin}(x_0)$ are present in \eq{e:caahqet}; the parameters
may thus be adjusted to absorb all infinities and with properly chosen coefficients the continuum
limit of the r.h.s. is expected to exist. The basic assumption of the effective field theory is that once the
finite parts of the coefficients have been 
determined by matching a set of observables to QCD, these coefficients are applicable
to any other observables. 

The B-meson mass is given by $\caa(x_0)$ in
large volume via
\bes
  \mB = - \lim_{x_0\to\infty} {\partial_0 + \partial_0^* \over 2}\,\log\caa(x_0)\,,
\ees
with
\bes
        \partial_0    f(x_0) =  {1 \over a} [f(x_0+a) - f(x_0)]\,,\quad 
        \partial_0^*  f(x_0) =  {1 \over a} [f(x_0) - f(x_0-a)]\,.
\ees
Inserting the HQET expansion we derive
\bes
    \mB = \mB^\mrm{stat} + \mB^{(1)} \,,
\ees
with 
\bes
    \mB^\mrm{stat}&=&\mhbare^\mrm{stat}+\Estat\,, \;  
    \Estat=  - \lim_{x_0\to\infty} {\partial_0 + \partial_0^* \over 2}\,\log\caa^\mrm{stat}(x_0) \,, 
    \label{e:estat}\\
    \mB^{(1)} &=& \mhbare^{(1)}+\omegakin \Ekin + \omegaspin \Espin\,, \\
    \Ekin &=& - \lim_{x_0\to\infty} {\partial_0 + \partial_0^* \over 2}\, \big[\caa^\mrm{kin}(x_0)/\caa^\mrm{stat}(x_0)\big]\,,\\
    \Espin &=& - \lim_{x_0\to\infty} {\partial_0 + \partial_0^* \over 2}\, \big[\caa^\mrm{spin}(x_0)/\caa^\mrm{stat}(x_0)\big]\,.
\ees
Here the terms $\propto \cahqet$ of \eq{e:caahqet}
do not contribute. They are proportional to the derivative
of ratios $\cdaa^\mrm{stat}(x_0)/\caa^\mrm{stat}(x_0)$. At large
$x_0$ these ratios approach a constant since $\delta \Astat$ has
the same quantum numbers as $\Astat$. 
Using the transfer matrix formalism 
(with normalization $\langle B | B \rangle = 1$), 
one further observes that
\bes
  \label{e:ekin}
    \Ekin = - \langle B | a^3\sum_{\vecz} \Okin(0,\vecz)| B \rangle_\mrm{stat} \,,\quad
  \label{e:espin}
    \Espin = - \langle B | a^3\sum_{\vecz} \Ospin(0,\vecz)| B \rangle_\mrm{stat} \,.
\ees
As expected, only the parameters of the action are relevant
in the expansion of a hadron mass. In the above relations
$\mhbare^\mrm{stat}$ absorbs a linear divergence of $\Estat$ 
and $\mhbare^{(1)}$ a quadratic divergence of $\Ekin$. 

Going through the same steps in the vector channel and
using the spin symmetry of the static action
is one way to see that the combination
\bes
   \label{e:mBav}
   {\mbav} \equiv {1\over 4} [\mB+3\mBstar] = \mhbare + \Estat +\omegakin \Ekin 
\ees
is independent of $\omegaspin$. It is instructive to represent 
this equation in a different way, subtracting the $1/a$ (and $1/a^2$) 
divergences of $\Estat$ (and $\Ekin$). In this way we have
\bes
  \label{e:mbavsplit}
  \mbav &=& \mb^{(0a)} +\mb^{(0b)} + \mb^{(1a)} +\mb^{(1b)}\,, \\
  \label{e:mb0a}
  \mb^{(0a)} &=& \mhbare^\mrm{stat}+E_\mrm{stat}^\mrm{sub}\,,\\
  \label{e:mb0b}
  \mb^{(0b)} &=& \Estat-E_\mrm{stat}^\mrm{sub}\,,\\
  \label{e:mb1a}
  \mb^{(1a)} &=& \mhbare^\mrm{(1)}+\omegakin\Ekin^\mrm{sub}\,,\\
  \label{e:mb1b}
  \mb^{(1b)} &=& \omegakin [\Ekin-\Ekin^\mrm{sub}]\,,
\ees
with finite terms $\mb^{(0a)},\mb^{(0b)},\mb^{(1a)},\mb^{(1b)}$. Our strategy,
described in the introduction can be seen as a way of determining 
the coefficient $\omegakin$ as well as the subtractions $\Estat^\mrm{sub},\Ekin^\mrm{sub}$
from finite volume computations in QCD and HQET. Finite parts in
the subtraction terms do of course depend on the detailed choice
of kinematical parameters such as the matching volume, but the end result
is unique up to terms of order $\minv^2$. Note that by the same logics, 
the order $\minv$ term, $\mb^{(1a)}+\mb^{(1b)}$, is {\em not unique} but depends
on the details of the strategy. 

Since the prediction \eq{e:mBav} 
requires only the knowledge of
two parameters, we also need only two 
finite volume observables to perform the matching with QCD. 
The \SF is particularly useful to find suitable
observables \cite{hqet:pap1,hqet:pap2,hqet:pap3}.
We proceed to discuss the $\minv$ expansion in this situation. 

\subsection{\SF correlation functions}

The pure gauge \SF has thoroughly been discussed in \cite{SF:LNWW},
relativistic and static quarks were introduced in \cite{SF:stefan1}
and \cite{zastat:pap1}. 
In particular in the last reference also
Symanzik $\Oa$-improvement was discussed. The improvement
of the \SF requires the addition of dimension four 
local composite fields localized at the boundaries~\cite{impr:pap1}. 
However, it turns out 
that there are no dimension four 
composite fields which involve static 
quarks fields and which are compatible with the symmetries of the 
static action and the \SF boundary conditions and which do not vanish by the
equations of motion. Thus there are no $\rmO(a)$ boundary counter terms with
static quark fields. For the same reason there are also no $\rmO(\minv)$ boundary
terms in HQET. 
This then means the HQET expansion of the boundary quark fields $\zeta,\bar\zeta$ 
is trivial up to and including $\minv$ terms. 

For details of the boundary conditions as well as the definition of the 
fields $\zeta,\bar\zeta$ we refer to \cite{zastat:pap1}, where also our notation
is explained. For a general understanding it is, however, sufficient 
to note a few facts. In space the fermion fields
are taken to be periodic up to a phase, 
\bes
  \psi(x+\hat{k}L) = \rme^{i\theta} \psi(x) \,, \quad \psibar(x+\hat{k}L) = \rme^{-i\theta} \psibar(x)\,,
\ees
with the same phase $\theta$ for all quark fields, whether relativistic
or described by HQET. In time we take homogeneous Dirichlet boundary 
conditions at $x_0=0$ and $x_0=T$ \cite{zastat:pap1}. Correlation functions
can be formed from composite fields in the bulk, $0 < x_0 < T$, and
boundary quark fields  $\zeta\,,\,\bar\zeta$.
In QCD, correlation functions in the pseudoscalar and vector channel are
\bes
  \fa(x_0,\theta) &=& -{a^6 \over 2}\sum_{\vecy,\vecz}\,
  \left\langle
  (\aimpr)_0(x)\,\zetabar_{\rm b}(\vecy)\gamma_5\zeta_{\rm l}(\vecz)
  \right\rangle  \,, \label{e_fa} \\
  \kv(x_0,\theta) &=& -{a^6 \over 6}\sum_{\vecy,\vecz,k}\,
  \left\langle
  (\vimpr)_k(x)\,\zetabar_{\rm b}(\vecy)\gamma_k\zeta_{\rm l}(\vecz)
  \right\rangle  \,. \label{e_kv}
\ees
The $\Oa$ improved currents $\aimpr,\vimpr$ can be found in \cite{hqet:pap1}.
Furthermore we consider boundary to boundary correlation functions
\bes
  \fone(\theta) &=&
  -{a^{12} \over 2L^6}\sum_{\vecu,\vecv,\vecy,\vecz}
  \left\langle
  \zetalbprime(\vecu)\gamma_5\zzetaprime_{\rm b}(\vecv)\,
  \zetabar_{\rm b}(\vecy)\gamma_5\zetal(\vecz)
  \right\rangle\,, \\
  \kone(\theta) &=&
  -{a^{12} \over 6L^6}\sum_{\vecu,\vecv,\vecy,\vecz,k}
  \left\langle
  \zetalbprime(\vecu)\gamma_k\zzetaprime_{\rm b}(\vecv)\,
  \zetabar_{\rm b}(\vecy)\gamma_k\zetal(\vecz)
  \right\rangle\,.
\ees
Their renormalization is standard~\cite{impr:pap5}, for example
\bes
   \left[\fa\right]_\mrm{R}(x_0,\theta) = \za \zzeta^2 \fa(x_0,\theta)\,, \quad
   \left[\fone\right]_\mrm{R}(\theta) = \zzeta^4 \fone(\theta)\,,
\ees
with $\zzeta$ a renormalization factor of the relativistic 
boundary quark fields.

In complete analogy to the case of 
a manifold without boundary we can write down the expansions to first order in $\minv$. 
They read
\bes
  \left[\fa\right]_\mrm{R} &=& \zahqet \zzetah\zzeta \rme^{-\mhbare x_0}
        \left\{ \fastat + \cahqet \fdeltaastat + \omegakin \fakin
                + \omegaspin \faspin
        \right\}\,, \label{e:faexp} \\
  \left[\kv\right]_\mrm{R} &=& \zvhqet  \zzetah\zzeta \rme^{-\mhbare x_0}
        \left\{ \kvstat + \cvhqet \kdeltavstat + \omegakin \kvkin
                + \omegaspin \kvspin
        \right\}\,,  \label{e:kvexp} \\ 
      &=& -\zvhqet  \zzetah\zzeta \rme^{-\mhbare x_0}
        \left\{ \fastat + \cvhqet \fdeltaastat + \omegakin \fakin
                -\frac13 \omegaspin \faspin
        \right\}\,, \nonumber \\
  \label{e:foneexp}
  \left[\fone\right]_\mrm{R} &=& \zzetah^2\zzeta^2 \rme^{-\mhbare T}
        \left\{ \fonestat + \omegakin \fonekin
                + \omegaspin \fonespin
        \right\}\,, \\
  \label{e:koneexp}
  \left[\kone\right]_\mrm{R} &=& \zzetah^2\zzeta^2 \rme^{-\mhbare T}
        \left\{ \fonestat + \omegakin \fonekin
                -\frac13 \omegaspin \fonespin
        \right\}\,.
\ees
Apart from
\bes
  \fdeltaastat(x_0,\theta) =  -{a^6 \over 2}\sum_{\vecy,\vecz}\,
  \left\langle
  \delta\Astat(x)\,\zetabar_{\rm h}(\vecy)\gamma_5\zeta_{\rm l}(\vecz)
  \right\rangle
\ees
the labeling of the different terms follows directly the one
introduced in \eq{e:exp}. We have used identities such as
$\fakin=-\kvkin\,,\; \faspin=3\kvspin$.
As a simple consequence of the spin symmetry of the static action,
these are valid at any lattice spacing.


\section{Finite volume observables and step scaling functions \label{s:phi}}

\subsection{Observables}

We concentrate on a strategy based on
the correlation functions $\fone,\kone$ alone.
This is advantageous, since the additional coefficients
$\cahqet,\cvhqet$ in  \eq{e:faexp},~\eq{e:kvexp} are avoided. 
Apart from the b-quark, we set the masses of all quarks to zero.

In terms of the spin-averaged combination,
\bes
  F_1(L,\theta) = {1\over 4} \big[\log \fone(\theta) +3 \log \kone(\theta)\big]\,,
\ees
we form 
\newcommand{\athalf}{\;\mbox{at}\;T=L/2}
\newcommand{\atxhalf}{\;\mbox{at}\;x_0=L/2\,,\;T=L}
\bes
        R_1(L,\theta_1,\theta_2) &=&  F_1(L,\theta_1) - F_1(L,\theta_2) \, \athalf \\
        \meffone(L,\theta_0) &=& -{\partial_T +  \partial_T^* \over 2}  F_1(L,\theta_0) \athalf\,.
\ees
Note that the boundary quark wave function renormalization cancels in $R_1$ and in $\meffone$.
They are thus finite after renormalization of the parameters of the Lagrangian.

The dimensionless observables,
\bes \label{e:defphi1}
       \Phi_1(L,\Mbeauty) &=& \ratone(L,\theta_1,\theta_2) - \ratonestat(L,\theta_1,\theta_2)\,, \\
       \Phi_2(L,\Mbeauty) &=& L\meffone(L,\theta_0) \,, \\
       \ratonestat(L,\theta_1,\theta_2) &=& \log \big[\fonestat(L,\theta_1)/\fonestat(L,\theta_2)\big]\,
        \athalf
\ees
are parametrized in terms of the RGI mass of the b-quark, $\Mbeauty$.
They have a particularly simple $\minv$ expansion
\bes \label{e:phi1exp}
        \Phi_1(L,\Mbeauty) &=& \omegakin \ratonekin(L,\theta_1,\theta_2)\,,\\
     \label{e:phi2exp}
        \Phi_2(L,\Mbeauty) &=& L\,\big[ \mhbare + \meffonestat(L,\theta_0) 
                               + \omegakin\meffonekin(L,\theta_0) \big]\,, 
\ees
which involves
\bes 
        \ratonekin(L,\theta_1,\theta_2) &=& {\fonekin(L,\theta_1)\over\fonestat(L,\theta_1)} 
        - {\fonekin(L,\theta_2)\over\fonestat(L,\theta_2)}\,  \athalf\,,\\
        \meffonestat(L,\theta_0) &=& -{\partial_T +  \partial_T^* \over 2}  \log\fonestat(\theta_0)\, 
                        \athalf\,, \label{e:meffone}\\
        \meffonekin(L,\theta_0)  &=& -{\partial_T +  \partial_T^* \over 2}  [\fonekin(\theta_0)/\fonestat(\theta_0)]\,
                        \athalf\,. \label{e:meffonekin}
\ees
The  $\theta_0,\theta_1,\theta_2$ dependence of $\Phi_i$ is not 
explicitly written, but will
of course be relevant in the numerical results. 
For the reader familiar with \cite{lat01:rainer,hqet:pap1}, we point out
that $\meffone$ differs from $\meff$ which was used in those 
references. 
Note that in \eq{e:defphi1} we subtract the static
term. This  simplifies 
subsequent formulae. 
In fact, whenever such a lowest order contribution is
universal (in the sense of having a universal continuum limit) 
and independent of an HQET parameter, it will be convenient to 
subtract it. 
Despite this subtraction, we refer to $\Phi_1$
as an observable in QCD.

The reader may be surprised that we introduce the quantity $\meffone$ which 
contains a (discretized) derivative with respect to the time extent, $T$.
Its MC evaluation requires two separate simulations~\footnote{
In \app{s:alt} we discuss a different strategy, which is based on
the $x_0$-derivative of $\fa$ and thus requires less simulations.
Note, however, that these additional simulations do not represent a
significant effort.
}.
However, obviously a quantity 
of order $\mbeauty$ is needed and this is obtained
from some logarithmic derivative of a correlation function. 
Boundary-to-boundary  
correlation functions are then very convenient since one does not have to
deal with the $\minv$ corrections to the currents. It is a useful feature 
of the \SF that such gauge invariant correlation functions are available.

\subsection{Step scaling functions \label{s:ssf}}

We turn to the relations between $\Phi_i(L,\Mbeauty)$ and
$\Phi_i(2L,\Mbeauty)$ in the effective theory. 
The dimensionful variable $L$ is replaced by
the \SF renormalized coupling $\gbar^2(L)$ \cite{alpha:SU3} over which
we have good control in numerical computations \cite{mbar:pap1}. 
Straightforward substitution yields
\bes
   \label{e:phi12l}
   \Phi_1(2L,\Mbeauty) &=& \sigmakin_1(u)\,  \Phi_1(L,\Mbeauty)\,,\\
   \label{e:phi22l}
   \Phi_2(2L,\Mbeauty) &=& 2\Phi_2(L,\Mbeauty)  + \sigmam(u) +  
        \sigmakin_2(u) \, \Phi_1(L,\Mbeauty)\,,
\ees
where always $u=\gbar^2(L)$. Our continuum step scaling functions
$\sigma$ (with any subscripts or superscripts) are defined in
terms of those at finite lattice spacing as
\bes
   \sigma(u) = \lim_{a/L \to 0} \Sigma(u,a/L)\,.
\ees 
At finite lattice spacing we have
\bes
   \Sigmakin_1(u,a/L) &=& \left.{\ratonekin(2L,\theta_1,\theta_2) \over
              \ratonekin(L,\theta_1,\theta_2) }\right|_{u=\gbar^2(L)}\,, \\
   \Sigmakin_2(u,a/L) &=&
                \left. {2L\,[\meffkin_1(2L,\theta_0) - \meffkin_1(L,\theta_0)]
                                        \over
                                  \ratonekin(L,\theta_1,\theta_2)  }\right|_{u=\gbar^2(L)} \,, \\
   \Sigmam(u,a/L) &=& 2L
                 \left[ \meffstat_1(2L,\theta_0) - \meffstat_1(L,\theta_0)
                                \right]_{u=\gbar^2(L)} \,.
\ees
The above equations are easily derived. 
In a first step, just from the $\minv$ expansions
of $\Phi_i$, one obtains them at a given resolution $a/L$
or equivalently at fixed bare coupling, $g_0$.
One then uses that $\Phi_i(L,\Mbeauty)$ are dimensionless 
physical observables with a continuum limit. 
Since the self energy of a static quark cancels in $\sigmam$,
also that quantity has a finite continuum limit. 
Thus the continuum limit
of the step scaling functions $\Sigmam,\Sigmakin_i$ exists 
and eqs.(\ref{e:phi12l},\ref{e:phi22l}) can be written in terms
of continuum quantities, as we have done.


\section{$\Mbeauty$ including $\minv$ corrections \label{s:mb}}

Before giving the equation for $\Mbeauty$, we recall the overall 
strategy. For $L_1\approx0.4\,\fm$ we compute 
$\Phi_1(L_1,\Mbeauty), \Phi_2(L_1,\Mbeauty)$ 
for a few quark masses around the physical one in quenched
QCD. 
It is understood that the continuum limit is reached by an extrapolation
and with a suitable interpolation of $\Phi_i$ in $\Mbeauty$,
these quantities can be considered to be known as a function
of $\Mbeauty$. 
With the step scaling functions described
in the previous section and computed in the effective theory,
we then  arrive at $\Phi_1(L_2,\Mbeauty), \Phi_2(L_2,\Mbeauty)$,
where $L_2=2L_1$. It remains to express the spin averaged 
B-meson mass 
${\mbav}$ in terms of $\Phi_1(L_2,\Mbeauty), \Phi_2(L_2,\Mbeauty)$.

To this end,
we straightforwardly combine 
eqs.~(\ref{e:phi1exp},\ref{e:phi2exp}) with \eq{e:mBav} and obtain
\bes
 L {\mbav} &=& \Phi_2(L,\Mbeauty) +
        L [E^\mrm{stat} - \meffstat_1(L,\theta_0)] +{L[E^\mrm{kin} - \meffkin_1(L,\theta_0)]
                 \over
                                                \ratonekin(L,\theta_1,\theta_2)} \Phi_1(L,\Mbeauty)\,.
        \nonumber \\
        \label{e:largel}
\ees
We now set $L=L_2$ in this equation and insert
\eq{e:phi22l}. In the form of \eq{e:mbavsplit} we then have
\bes
\label{e:0a}  
   L_2\mB^\mrm{(0a)}(\Mbeauty) &=&  \sigmam(u_1) + 2\,\Phi_2(L_1,\Mbeauty) \,\\
\label{e:0b}  
   L_2\mB^\mrm{(0b)} &=& L_2 [E^\mrm{stat} - \meffstat_1(L_2,\theta_0)]\,,\\
\label{e:1a}  
 L_2\mb^{(1a)}(\Mbeauty) &=& \sigmakin_2(u_1) \,\Phi_1(L_1,\Mbeauty)\,,
               \;  \\
\label{e:1b}  
 L_2\mb^{(1b)}(\Mbeauty) &=&  L_2{E^\mrm{kin} - \meffkin_1(L_2,\theta_0) \over \ratonekin(L_2,\theta_1,\theta_2)}
         \,\sigmakin_1(u_1)\, \Phi_1(L_1,\Mbeauty) \,, 
  \label{e:master}
\ees
where 
\bes
        u_1=\gbar^2(L_1)\,,\quad L_2=2L_1\,.
\ees
The subtraction of power divergences in \eq{e:mb0b}, \eq{e:mb1b}
are 
$E_\mrm{stat}^\mrm{sub}=\meffstat_1(L_2,\theta_0)$, $\Ekin^\mrm{sub}=\meffkin_1(L_2,\theta_0)$
and  $\sigmakin_1(u_1)\, \Phi_1(L_1,\Mbeauty) / \ratonekin(L_2,\theta_1,\theta_2)$ is 
a representation of 
the bare parameter $\omegakin$ in \eq{e:mb1b}. The other parts, $\mb^{(0a)},\mb^{(1a)}$,
are computable entirely in finite volume. 

The step scaling functions $\sigma$ have been discussed before. 
They can be computed with lattice spacings such that $a/L_1$
is reasonably small, say below $1/6$. Of course they should be 
extrapolated to the continuum. We work with lattice spacings $a \leq 0.07 \,\fm$
in this step. 
The relativistic observables $\Phi_i(L_1,\Mbeauty)\,,\,i=1,2$ 
are computed for $a \leq 0.02 \,\fm$, where a relativistic b-quark can  
be described by the $\Oa$-improved Wilson action with
controlled $a^2$-effects. Finally,
the combinations $ L_2 [E^\mrm{stat} - \meffstat_1(L_2,\theta_0)]$ and
$L_2{E^\mrm{kin} - \meffkin_1(L_2,\theta_0) \over \ratonekin(L_2,\theta_1,\theta_2)}$ are computed
for lattice spacings of $a\leq0.1\,\fm$ such that finite size effects 
in $\Estat$ and $\Ekin$ 
are negligible on lattices with an affordable number of points.

The mass of the b-quark is obtained from \eq{e:mbavsplit} by expanding
\bes
   \Mbeauty = \Mbeauty^{(0)} +  \Mbeauty^{(1)} \,,
\ees
where $\Mbeauty^{(0)}$ is the solution of the static equation
\bes
\label{e:mbeq0}
   \mbav = \mB^\mrm{(0a)}(\Mbeauty^{(0)}) + \mB^\mrm{(0b)}(\Mbeauty^{(0)})
\ees
and the $\minv$ correction 
is
\bes
   \Mbeauty^{(1)}&=&-{1\over S}\big[\mB^\mrm{(1a)}(\Mbeauty^{(0)}) + \mB^\mrm{(1b)}(\Mbeauty^{(0)})\big]
\ees
with
\bes
   S &=& 
{\rmd \over \rmd \Mbeauty} \big[\mB^\mrm{(0a)}(\Mbeauty) + \mB^\mrm{(0b)}(\Mbeauty)\big]
      = {\rmd \over \rmd \Mbeauty} [\mB^\mrm{(0a)}(\Mbeauty)  \big] \,.
\ees

We finish the discussion of the strategy with a remark on the 
dependence on the mass of the light quarks. This is relevant 
because it is of course better to consider the spin-averaged 
$B_\mrm{s}$ quark mass in \eq{e:mBav};  the necessary large volume 
computations are easier than
for the $B_\mrm{d}$ meson. In the quenched approximation the parameters 
in the action $\mhbare,\omegakin$ are independent of the light
quark mass.\footnote{
In general, 
$\dmstat$ (and hence also $\mhbare$)
will contain a term like $b(g_0)m_\mrm{l}$,
where for simplicity the light quarks are assumed to be degenerate
with mass $m_\mrm{l}$. Obviously, 
$b(g_0) =\rmO(g_0^4)$ does, however, vanish for $\nf=0$.
As a renormalization term odd in $m_\mrm{l}$,
it is also absent for twisted mass lattice QCD \cite{tmqcd:pap1}
and QCD with exact chiral 
symmetry
\cite{exactchi:neub,exactchi:martin,exactchi:perfect,exactchi:shamir}.
}
Since our strategy determines them through 
finite volume computations, it follows that in all these 
computations the light quark mass may be set to zero, a 
convenient choice. Only  $\Ekin$ and $\Estat$ are then
to be computed at the mass of the light quark of the meson
who's (spin averaged) mass is considered.


\section{Results \label{s:res}}

\FIGURE{
\includegraphics*[width=6.9cm]{plots/Phi2}
\hspace{0.5cm}
\includegraphics*[width=6.9cm]{plots/Phi3}
\caption{\footnotesize
Continuum extrapolation of $\Phi_2(L_1,\Mbeauty)$, 
for $z=10.4\,,\; 12.1\,,\; 13.3$ from bottom to top. The errors
in the relation between bare quark mass $\mqtilde$ and
the RGI mass $M$ are translated into errors in
$\Phi_2$. The $g_0$--independent part of that error
is included {\em after}~\cite{hqet:pap2} the continuum
extrapolation (left side error bar).
On the right, the equivalent in the alternative strategy
is shown for $\theta_0=1/2$ (see \app{s:alt}).
}\label{f:contphi2}
}

We have performed a numerical computation in the 
quenched approximation, using the $\Oa$ improved
Wilson action \cite{impr:SW,impr:pap1,impr:pap3}. 
The box size $L_2$ is chosen as $L_2=1.436r_0$,
where $r_0$,  defined in terms of the 
static quark potential~\cite{pot:r0}, has a 
phenomenological value of $r_0\approx0.5\,\fm$. 
From \cite{mbar:pap1} we know the \SF coupling 
$\gbar^2(L_1)=\gbar^2(L_2/2)\approx3.48$. 
Given the knowledge of $r_0/a$ as a function
of $g_0$ of Ref.~\cite{pot:r0_SU3} and  that of
the renormalized coupling \cite{mbar:pap1}, it
is then convenient to fix 
$g_0$ in different ways for the different
steps of the calculation. The differences
are of course only $a$-effects which disappear
in the continuum extrapolations.  We give more 
details below. 
We will take the uncertainties
in the relations $\gbar^2(L_1)\approx3.48$
and $\gbar^2(L_1/4)\approx1.8811$ (which we need later)
into account in the very end.

In order to complete our
definitions, we further choose $\theta_0=0$ and $\theta_1,\theta_2\in \{0,1/2,1\}$.
The different values of $\theta_1,\theta_2$ offer the
possibility to check whether our final results are independent
of these arbitrary parameters as they should
be up to small $\minv^2$ terms. Simulation parameters as well as raw results
are listed in tables in \app{s:sim} and \ref{s:large}.

\subsection{QCD observables\label{s:res1}}

For this part of the computation, we determined the
 bare parameters as in \cite{hqet:pap2}: 
$g_0$ is fixed by requiring
$\gbar^2(L_1/4)=1.8811$ for given resolutions $a/L$.
The PCAC mass of the light quark, defined exactly
as in that reference, is set to zero. 
Our heavy quark masses are chosen such that
$z=\Mbeauty\,L_1 \approx 10-13 $\,.
The bare parameters are listed in \tab{t:qcdpar}. 

\FIGURE{
\includegraphics*[width=6.9cm]{plots/R1}
\hspace{0.5cm}
\includegraphics*[width=6.9cm]{plots/R1_stat}
\caption{\footnotesize
Continuum extrapolation of $\Phi_1(L_1,\Mbeauty)$,
separately for $\ratone(L_1,1/2,1)$ in QCD (left) and
for  $\ratonestat(L_1,1/2,1)$ in the static approximation (right).
Circles denote results with action HYP1 and squares, 
displaced slightly for visibility, are from action HYP2.
The corresponding continuum extrapolation lines are
slightly displaced as well.
}\label{f:contphi1}
}

We focus our attention directly on the continuum extrapolations.
As an example we show $\Phi_1(\Mbeauty,L_1)$ and $\Phi_2(\Mbeauty,L_1)$ in 
\fig{f:contphi1} and \fig{f:contphi2}. 
Note that for the static subtraction $\ratonestat(L_1,1/2,1)$,
displayed on the right of \fig{f:contphi1}, 
our lattice spacings are 
roughly a factor three larger, since in the effective
theory we only have to respect $a/L_1 \ll 1$, not
$a\Mbeauty \ll 1$ (for details see \app{s:sim}). 
Data have been obtained
for two static actions, $\hypone$ and \hyptwo \cite{stat:actpaper}.
In fitting them to the expected
$a$-dependence, their continuum limit value
is constrained to be independent of the action,
but the $a^2$ slopes are of course different.
The data for the different actions are highly correlated. 
As in all such cases, the errors of the continuum limit 
are computed from jacknife samples.

For values of $\theta_i$ which differ from the
choice made in the figures, the $a$-dependence 
is very similar. In all these cases we find that
extrapolations linear in $a^2$ using all four available
lattice spacings are compatible with the ones 
where the data point at largest lattice spacing is 
ignored. 
We take the extrapolations with three points
as our results for further analysis, since they have the
more conservative error bars. The continuum limits are listed 
together with the raw numbers 
in Tables \ref{t:R1stat} and \ref{t:phi12}.  From a fit of 
the continuum $\Phi_2(z)$ to a linear function,
we then extract the slope
\bes
  S = {\rmd\over\rmd z} \Phi_2 = 0.61(5)\, \label{e:Svalue}
\ees
and we are done with the matching. The rest of 
the numerical computations is carried out in the effective theory.

\subsection{HQET step scaling functions\label{s:res2}}

\FIGURET{
\includegraphics*[width=6.5cm]{plots/sigmam}
\caption{\footnotesize
Continuum extrapolation of $\Sigmam$ 
and $\widetilde\Sigmam$. 
}\label{f:sigmam}
}
Next we discuss the connection of  $\Phi_i(L_1,\Mbeauty)$ to 
$\Phi_i(L_2,\Mbeauty)$, $L_2=2L_1$. It is given 
by the step scaling functions of \sect{s:ssf}.
The bare parameters used in their computation 
are described in \app{s:sim}, 
and the values at finite resolution $a/L_1$ are given in
Tables \ref{t:Sigmam}-\ref{t:Sigmakin2}~\footnote{
For $\Sigmam,\Sigmakin_2$ the coarsest resolution considered is
$a/L_1=8$. Due to the derivative $\partial_T$
at $T=L/2$, smaller values of $L_1/a$ would involve 
a very short time separation.}.

\FIGUREP{
\includegraphics*[width=6.5cm]{plots/Sigma1kin}
\caption{\footnotesize
Continuum extrapolation of $\Sigmakin_1$ for $\theta_1=1/2,\theta_2=1$. 
}\label{f:sigmakin1}
}

At lowest order in $\minv$, only $\sigmam$ contributes.
In its continuum extrapolation (\fig{f:sigmam}, \tab{t:Sigmam})
we allow for a slope in $a^2$, although the data are compatible 
with a vanishing slope. 
Note that the {\em absolute} error of $\sigmam$ is negligible 
in comparison to twice the one of
$\Phi_2$ (see \fig{f:contphi2}) to which it
is added in \eq{e:0a}. In fact the uncertainty
in $\sigmam$ corresponds to an error
of only $5\,\MeV$ in the b-quark mass, illustrating the
possible precision in the static effective theory
with these actions \cite{stat:actpaper,stat:letter}.

\FIGUREP{
\includegraphics*[width=6.5cm]{plots/mB1a_2b2}
\hspace{0.5cm}
\includegraphics*[width=6.5cm]{plots/mB1a_3b3}
\caption{\footnotesize
Lattice spacing dependence of $L_2 \mb^{(1a)}$ for $\Mb=\Mb^{(0)}$. 
On the left
we show $\mb^{(1a)}$ as introduced in \sect{s:mb}, 
with $\theta_1=1/2,\; \theta_2=1$.
We insert $\Phi_1$ in the continuum limit, such that
the lattice spacing dependence is just due to $\Sigmakin_2$.
On the right the corresponding quantity is shown for the
alternative strategy of \app{s:alt}, again with continuum
values for $\widetilde\Phi_i$. There we set 
$\theta_0=1/2\;,\theta_1=1/2\;,\theta_2=1$. 
}\label{f:mb1a}
}

A relevant question is how the precision 
deteriorates when one includes the first order corrections in $\minv$.
Then two more step scaling functions contribute. In \fig{f:sigmakin1},
we illustrate how
the continuum limit of 
$\sigmakin_1$ is obtained. Here
we have to allow for a {\em linear} dependence on
the lattice spacing, since the theory is not $\Oa$ improved
at the level of the $\minv$ contributions \cite{hqet:pap1}. 
Taking the more conservative 
fit with only three points, we arrive at the continuum limit
listed in \tab{t:Sigmakin1} for all combinations $\theta_1,\theta_2$.
In \eq{e:1b}, $\sigmakin_1$ is multiplied by small numbers (of order 
$\minv$). This means that its error will be negligible
in the overall error budget. 

Instead of $\sigmakin_2$ we show directly the continuum extrapolation
of $\mb^{(1a)}$, \eq{e:1a}. 
As for 
$\Sigmam$, the data shows no significant
$a$--dependence.  Nevertheless, in order to have a realistic
error estimate, we allow for
a linear slope in $a$ (\fig{f:mb1a}). In \tab{t:Sigmakin2} 
we list the raw numbers for $\Sigmakin_2$ as well as the
extracted continuum limit for further analysis.

\subsection{Large volume matrix elements and $\Mbeauty$\label{s:res3}}

\FIGURE{
\includegraphics*[width=6.9cm]{plots/mB0b_2b2}
\hspace{0.5cm}
\includegraphics*[width=6.9cm]{plots/mB0b_3b3}
\caption{\footnotesize
Lattice spacing dependence of $L_2 \mb^{(0b)}$, details
as in  \fig{f:mb1a}. 
}\label{f:mb0b}
}

\FIGURE{
\includegraphics*[width=6.9cm]{plots/mB1b_2b2}
\hspace{0.5cm}
\includegraphics*[width=6.9cm]{plots/mB1b_3b3}
\caption{\footnotesize
Lattice spacing dependence of $L_2 \mb^{(1b)}$, details
as in  \fig{f:mb1a}.
}\label{f:mb1b}
}

The last missing pieces in \eq{e:mbavsplit} are the large volume
static energy $\Estat$, \eq{e:estat}, and the matrix element of the kinetic
operator $\Ekin$, \eq{e:ekin}. 
Here, in contrast to the rest of our numerical evaluations, 
the light quark mass is set to the mass of the strange quark
in order to avoid a chiral extrapolation. The spin averaged 
mass of the $B_\mrm{s}$ system is then to be inserted into \eq{e:mBav}.

Although $\Estat$ and $\Ekin$ can be computed
with periodic boundary conditions we here follow 
\cite{mbar:pap2} and evaluate also these quantities with \SF 
boundary conditions in a large volume of about $T\times(1.5\fm)^3$,
with $1.5\fm \leq T\leq 3\fm$ 
(also a check for finite size effects is carried out). 
The extraction is fairly standard,
but still care has to be taken to make sure that the 
ground state contribution is obtained. This is a particularly
relevant issue for B-physics, because the gap to the first 
excited state is rather small. We relegate details to \app{s:large}
and discuss immediately the universal combinations $L_2 \mb^{(0b)}$,
$L_2 \mb^{(1b)}$ which enter in \eq{e:mbavsplit}.
The static contribution, shown in \fig{f:mb0b}, 
is known with very good precision~\footnote{
We show the results given for the static action 
HYP2. The continuum extrapolation with action HYP1
looks very similar, but the fit has a smaller $\chi^2$.
}.
In contrast, the $\minv$ correction
$L_2 \mb^{(1b)}$ does have a noticeable total uncertainty
(\fig{f:mb1b}, \tab{t:Mb_2b2}). 
Still, this error is only about 50\% of the one on $2\Phi_2$.
Note also that this error is almost entirely due to
$\Ekin$ which may possibly be computed more precisely by
other techniques \cite{michael:alltoall}. 

We now have all pieces necessary to solve the equations for $\Mbeauty$. 
The static one, \eq{e:mbeq0}, is illustrated in \fig{f:staticmb}.
The horizontal error band is given by subtracting 
the static pieces $\sigmam+L_2\mb^{(0b)}$ from the  experimental
number 
\bes
   \mbav= 5.405\,\GeV\,.
\ees 
The figure demonstrates again that the main source of error is
contained in the QCD computation of $\Phi_2$.
Finally, by interpolating $\Phi_i(L_1,\Mbeauty)$ to 
$\Mbeauty=\Mbeauty^{(0)}$ we obtain ($\theta_1=1/2,\theta_2=1$)
\bes
   r_0\,\Mbeauty^{(0)} &=& 17.25(20)\,,\; \Mbeauty^{(0)}=  6.806(79)\, \GeV\quad\mbox{for}\;r_0=0.5\,\fm
 \label{e:Mb0}\\
   r_0\,\Mbeauty^{(1)} &=& -0.12(9) \,,\; \Mbeauty^{(1)}= -0.049(39)\, \GeV\quad\mbox{for}\;r_0=0.5\,\fm\,
  \label{e:Mb1} \\
  r_0\,\Mbeauty &=& 17.12(22)\,,\; \Mbeauty =  6.758(86)\, \GeV\quad\mbox{for}\;r_0=0.5\,\fm\,.
  \label{e:Mbres} 
\ees
Here the small difference $\gbar^2(L_1/4)-1.8811$ as well as the 
statistical uncertainties in $\gbar^2(L_1)$ and $L_1/r_0$ have been taken into
account, as explained in \app{s:shift}.
Moreover, one can see in \tab{t:Mb_2b2} that the  $\theta_i$ dependence of the $\minv$ 
contribution is absorbed. 
\TABLE{
\begin{tabular}{ccccc}
\hline\hline \\[-1.75ex]
$\theta_1$ & $\theta_2$&&  $r_0\Mb^{(1a)}$ & $r_0\Mb^{(1b)}$ \\[0.2ex]
\hline\hline
0   & 1/2 && -0.06(3) & -0.06(8) \\
1/2 & 1   && -0.06(3) & -0.06(8) \\
1   & 0   && -0.06(3) & -0.06(8) \\
\hline\hline
\end{tabular}
\caption{\footnotesize
RGI results of $\minv$ correction of the b-quark mass, in
units of $r_0$.
}\label{t:Mb_2b2}


}
With $\Lambda_\msbar r_0=0.602(48)$~\cite{alpha:SU3,pot:intermed},
the 4-loop $\beta$ function and the mass anomalous 
dimension~\cite{MS:4loop1,MS:4loop2,MS:4loop3,MS:4loop4},
we translate 
$\Mbeauty= \Mbeauty^{(0)}+\Mbeauty^{(1)}$
to the mass in the $\msbar$ scheme at its own scale,
\bes
  \mbar_\beauty(\mbar_\beauty)= 4.347(48) \GeV\,;
  \label{e:msbarmass}
\ees
the associated perturbative uncertainty can safely be neglected. 
For completeness we note that in the $\msbar$ scheme the $\minv$ term amounts to $ -27(22)\MeV$.
\FIGURE{
\includegraphics*[width=6.9cm]{plots/Interp_2b2}
\hspace{0.5cm}
\includegraphics*[width=6.9cm]{plots/Interp_3b3}
\caption{\footnotesize
Graphical solution of \eq{e:mbeq0}. On the left hand side,
data points are $2\Phi_2$ and
the horizontal error band is $ L_2 {\mbav} - \sigmam -  L_2 [E^\mrm{stat} - \meffstat_1(L_2,\theta_0)]$.
On the right hand side, the analogous terms are shown for the alternative
strategy ($\theta_0=1/2$, see \sect{s:res4}).
}\label{f:staticmb}
}

\subsection{Comparison to results from an alternative strategy\label{s:res4}}

As mentioned earlier, at first sight it appears more natural 
to base the computation of $\Mbeauty$ on the logarithmic
derivative of the spin average of $\fa$ and $\kv$ as the prime finite volume quantity.
We have not chosen this option as our standard strategy
since then three observables are needed for matching. However,
it is useful to consider also that alternative strategy in order to
perform an explicit check that $\minv^2$ terms are as small as
expected. The results can be appreciated without detailed
definitions of the observables and step scaling functions,
the interested reader can find them in \app{s:alt}.
Here we note that
within this alternative strategy we actually  have
nine different sets of $\{\widetilde\Phi_1,\widetilde\Phi_2,\widetilde\Phi_3\}$. Only 
one observable, $\widetilde\Phi_1=\Phi_1$, is in common to the 
two strategies. For our graphs we have selected (arbitrarily) one choice 
of parameters.

\TABLE{
\begin{tabular}{ccccccccccccc}
\hline\hline \\[-1.75ex]
$\theta_1$ & $\theta_2$ & \multicolumn{6}{c}{$r_0\,(\Mb^{(0)} + \Mb^{(1a)} + \Mb^{(1b)})$} 
\\[0.2ex]
\hline\\[-1.75ex]
&  &&  Main strategy &&  \multicolumn{3}{c}{Alternative strategy} \\
 && && & $\theta_0 = 0$ &  $\theta_0 = 1/2 $ & $\theta_0 = 1$ \\
\hline
0   & 1/2  && 17.12(22)  && 17.25(28) &  17.23(28) & 17.17(32)  \\
1/2 &  1   && 17.12(22)  && 17.23(27) &  17.21(27) & 17.14(30)  \\
1   & 0    && 17.12(22)  && 17.24(27) &  17.22(28) & 17.15(30)  \\
\hline\hline
\end{tabular}
\caption{\footnotesize
RGI results of $\Mb$ inlcuding the $\minv$ correction, and 
comparison of the two strategies.
}\label{t:Mb_sum}


}

First, let us summarize what kind of differences one expects in such a
comparison apart from $a$-effects. In the order of magnitude counting,
we take $L_1^{-1} \sim \Lambda \sim 0.5\,\GeV$ and of course $L_2=2L_1$.
The matching observables $\meffone,\meffav$
are constructed to be
equal to the quark mass at the 
leading order in the HQET expansion.
They start to differ 
at the next to leading order, which means
by terms of order $\Lambda$. Also their 
dependence on $\theta_i$
is of that magnitude. Since $\Phi_2(L_1,\Mbeauty)$ 
and $\tilde\Phi_3(L_1,\Mbeauty)$ have been 
made dimensionless by multiplication with $L_1$
and $L_1$ happens to be around $\Lambda^{-1}$, the
differences of $\Phi_2(L_1,\Mbeauty)$ 
and $\tilde\Phi_3(L_1,\Mbeauty)$ are order one.
The step scaling functions $\sigmam,\widetilde\sigmam$ as well as
$L_2\mb^{(0b)}$ are added to $\Phi_2$ (or $\tilde\Phi_3$) to obtain
$L_2\mb$ in static approximation. Thus they depend on the details
at the same level, apart from a trivial $L_2/L_1 = 2$ factor. 
Of course, in the total static
estimate $r_0 \Mbeauty^{(0)}$ this dependence is reduced to
$r_0 \Mbeauty \times (\Lambda/\mbeauty)^2 \sim 1/5$. In the same way,
the $\minv$ corrections $L_2\mb^{(1a)},L_2\mb^{(1b)}$ themselves 
have a dependence on the
matching conditions which is $L_2 \times\Lambda^2/\mbeauty \sim 1/5$ but
the final result $r_0 \Mbeauty$ including these
terms is accurate and unique up to
$r_0 \times(\Lambda^3/\mbeauty)^2 \sim 1/50$ corrections.

We leave it to the reader to check in Fig.~\ref{f:contphi2} to \ref{f:mb1b}
that these expectations are fully satisfied by our results~\footnote{ 
We note in 
passing that $\widetilde \Sigmam$, in contrast to $\Sigmam$, does in principle
require an improvement coefficient, $\castat$ \cite{zastat:pap1}, for
$\Oa$-improvement. It has been set to the 1-loop values from \cite{stat:actpaper},
but the results are rather insensitive to $\castat$, so its uncertainty can be neglected.
}.
In fact it appears that our estimate for the expansion 
parameter, $\Lambda/\mbeauty \sim 1/10$ is quite realistic.
Of course, to find this out requires an explicit computation of the 
correction terms as presented here.
In some cases, such as 
 $\mb^{(1b)}$, our 
precision is not good enough to resolve a dependence on the 
matching conditions.

In the b-quark mass in the static approximation, $r_0 \Mbeauty^{(0)}$ 
(\eq{e:Mb0} and \tab{t:Mb_3b3}), 
the maximum difference is $0.5(2)$, which is of the predicted order of magnitude.
Finally, when we add all contributions together, the results
from the alternative strategy, \tab{t:Mb_sum}, are fully in agreement 
with \eq{e:Mbres}. As expected $\minv^2$ terms are not visible with our precision.
They can safely be neglected.


\section{Conclusions \label{s:concl}}

The main conclusion of this work is that fully non-perturbative
computations in lattice HQET, as they have been suggested in
\cite{hqet:pap1}, are possible in practice.
In particular, the uncertainties in the $\minv$ corrections are 
smaller than those in the static approximation, 
despite the fact that we
numerically cancel large $a^{-2}$ divergences in the $\minv$ terms.
The final error in the mass of the b-quark is
dominated by the uncertainty in the renormalization in QCD. 
Errors due to simulations in the effective theory can almost
be neglected in comparison. 

A very nice result is the independence of the final numbers 
for $\Mbeauty$ of
the matching condition: \tab{t:Mb_sum}  shows that 
within our reasonably small uncertainties, we get the same 
results for the quark mass for altogether twelve different 
matching conditions. This is expected up to very small terms
of order $r_0\Mbeauty \times (\Lambda/\mbeauty)^3 \sim 0.02$,
which should be
compared to our result $r_0\Mbeauty = 17.12(22) - 17.25(28)$.
Here the quoted range is due to the different matching conditions.
In the order of magnitude estimates we have made 
a guess for the typical scale of $\Lambda\sim0.5\,\GeV$. 
In the static approximation, some
of the matching conditions yield slightly differing results for 
the quark mass in agreement with the expectation for
such variations of $r_0\Mbeauty \times (\Lambda/\mbeauty)^2 \sim 0.2$. 

Both this explicit test of the magnitude of the different orders 
in the expansion and the naive order of magnitude estimate 
say that $\minv^2$ corrections are completely negligible.  

Still, in aspects of the computation, considerable improvement
can be envisaged.
For example, return to the $\minv$ contribution to the B-meson mass
\fig{f:mb1b}. The statistical errors grow rapidly as one decreases 
the lattice spacing. The by far dominating uncertainty in the 
shown combination is the
one of the large volume matrix $\Ekin$. It appears worth while 
to look for improvements, maybe along the line of 
\cite{alltoall:dublin}.
Due to these errors, and of course the missing $\Oa$-improvement of
the theory at order $\minv$ \cite{hqet:pap1}, 
the continuum extrapolation is not easy.
Fortunately it is still precise enough for the present case. It will be very 
interesting to see cases where the $\minv$ 
corrections are larger, as it is expected, for example, for $\fb$. 

Let us now turn to the computed value of $\mbar_\beauty$, \eq{e:msbarmass}. 
Starting from a precisely specified input, namely $r_0$, $\mk$ and
$(\mbs +3 \mbsstar)/4$, the value of $\Mbeauty$ is unambiguous
in the quenched approximation,
because these inputs fix the bare coupling, strange and beauty quark masses.
We have used the experimental meson masses and $r_0=0.5\,\fm$. 
Our numbers for $\Mbeauty$ or $\mbar_\beauty$ may then
be used as a benchmark result
for other methods. Indeed, a comparison shows agreement with
\cite{romeII:mb} and 
the recent extension of that work~\cite{lat06:damiano}
$\mbar_\beauty = 4.42(7)\,\GeV$. 

Earlier, the review \cite{lat00:lubicz} quoted 
$\mbar_\beauty=4.30(5)(5)\,\GeV$ and $\mbar_\beauty=4.34(3)(6)\,\GeV$, 
based on static computations~\cite{mb:Gimenez2} and an extrapolation
of NRQCD results to the static limit~\cite{mb:NRQCD} respectively. 
A perturbative subtraction
\cite{mbstat:MaSa,mb:dmstat_Direnzo,mb:dmstat_Lepage} 
of the linear divergence $\dmstat$ was carried out
in these static estimates and, of course, a continuum extrapolation
could not be done. 

However, if other inputs are used, the result may change because $r_0$ 
is only approximately known and because the quenched approximation is not 
real QCD. A rough idea on the possible changes can be obtained by varying
$r_0$ by $\pm 0.05\,\fm$. This changes $\mbar_\beauty(\mbar_\beauty)$
by roughly $\, \pm 80 \,\MeV$. 

These remarks just serve to stress the obvious necessity 
of performing computations with $\nf>0$. The ALPHA-collaboration
is  presently starting
with $\nf=2$, where the renormalization of the quark mass in QCD is
known \cite{mbar:nf2}. The necessary HQET computations are {\em not}
expected to be a big numerical challenge, apart from the large 
volume B-meson matrix elements: simulations of
the \SF for $L\leq1\,\fm$ are not very demanding with 
nowadays computing capabilities~\cite{algo:GHMCalpha}. 
Altogether the extension of the present work to
full QCD is feasible and should be carried out, since
presently no better method is known to compute the 
b-quark mass from lattice QCD.


\vspace{0.4cm}
{\bf Acknowledgements.} 
We thank Stephan D\"urr for collaboration in the early stages
of this work \cite{lat04:stephan}.
We thank NIC  for allocating computer time on the APEmille
computers at DESY Zeuthen to this project and the APE group for its help. This
work is supported by the  Deutsche Forschungsgemeinschaft 
in the SFB/TR 09.

\begin{appendix}

\section{Finite volume simulations \label{s:sim}}

\TABLE{
\begin{tabular}{cccccccc}
\hline\hline \\[-1.75ex]
$L\over a$   &   $\beta$  &  $\kappa_{\rm l}$  & $\bar{g}^2({L\over4})$  & $Z_{\rm P}(g_0,{L\over2})$ & $b_{\rm m}$
& $Z$  & $\kappa_{\rm h}$ \\[1ex]
\hline\hline \\[-1.75ex]
$20$ & $7.2611$ & $0.134145$ & $1.8811(19)$ & $0.6826(3)$ & $-0.621$ & $1.0955$ & $0.124195$ \\
     &          &               &              &             &               &             & $0.122119$ \\
     &          &               &              &             &               &             & $0.120483$ \\
$24$ & $7.4082$ & $0.133961$ & $1.8811(22)$ & $0.6764(6)$ & $-0.622$ & $1.0941$ & $0.126055$ \\
     &          &               &              &             &               &             & $0.124528$ \\
     &          &               &              &             &               &             & $0.123383$ \\
$32$ & $7.6547$ & $0.133632$ & $1.8811(28)$ & $0.6713(8)$ & $-0.622$ & $1.0916$ & $0.127991$ \\
     &          &               &              &             &               &             & $0.126967$ \\
     &          &               &              &             &               &             & $0.126222$ \\
$40$ & $7.8439$ & $0.133373$ & $1.8811(22)$ & $0.6679(8)$ & $-0.623$ & $1.0900$ & $0.128989$ \\
     &          &               &              &             &               &             & $0.128214$ \\
     &          &               &              &             &               &             & $0.127656$ \\
\hline\hline
\end{tabular}
\caption{Bare parameters used in the computation of
the QCD observables for $L=\tilde L_1$. 
}\label{t:qcdpar}

}

For the matching in a finite volume, 
we performed one set of simulations of
(quenched) QCD and one of HQET. 
In the case of the relativistic theory, we used $L=\tilde{L}_1$,
defined by  $\bar{g}^2(\tilde{L}_1/4)=1.8811$~\footnote{$\tilde{L}_1$ 
differs slightly from $L_1$ defined in the main text 
by $L_1=0.718 r_0$.
This mismatch is however corrected, as explained 
later in this appendix and in \app{s:shift}.}.
The parameters of these simulations 
have been taken from~\cite{hqet:pap2} (see \tab{t:qcdpar}). The difference is that
here $L=\tilde{L}_1=2\,\tilde{L}_0$ (and $T=L/2$ and $T=L/2 \pm a$ in
addition to $T=L$) compared to $L=\tilde{L}_0$ in \cite{hqet:pap2}.

The parameters for the resolution $\tilde{L}_1/a=20$ cannot be found in the
mentioned reference. For that point, the gauge coupling $\beta$ 
has been chosen such that $\bar{g}^2(\tilde{L}_1/4)=1.8811$ 
for $\tilde{L}_1/4a=5$, see~\cite{mbar:pap1}.  
The renormalization constant $\zp$ and 
$\kappa_{\rm l}=\kappa_{\rm c}$ have been computed here, while
$b_{\rm m}$ and $Z$ have been extrapolated from the values in Table~2 of~\cite{hqet:pap2}. 
These factors are put into
the relationship between the bare mass $m_{\rm q,h}$
and the RGI mass \cite{mbar:pap1,impr:babp},
\begin{equation}
  M=h \, Z_{\rm m} \, m_{\rm q,h} \left(1+b_{\rm m} am_{\rm
    q,h}\right)\,,
\end{equation}
where
\begin{equation}
Z_{\rm m} ={{Z \; Z_{\rm A}}\over{Z_{\rm P}}} \;, \quad {\rm and} \quad 
am_{\rm q,h}={{1}\over{2}}\left({{1}\over{\kappa_{\rm
      h}}}-{{1}\over{\kappa_{\rm c}}}\right) \;.
\end{equation}
The renormalization constant $Z_{\rm A}(g_0^2)$ is known non-perturbatively
from~\cite{impr:pap4}, while 
\begin{equation}
\label{h_factor}
h={{M}\over{\overline{m}(\mu_0)}}=1.544(14)\;, \;\; \mu_0=2/\tilde{L}_1 \;,
\end{equation}
relates the running quark mass in the Schr\"odinger functional 
scheme~\cite{mbar:pap1}
at the scale $\mu_0$, 
to the renormalization group invariant quark mass $M$~\footnote{
In $h={{M}/{\overline{m}(\mu_0)}}$ we take the small difference 
between the above defined $\tilde{L}_0$ and the value $L_0=L_1/2$ 
into account. It causes a change of less than 1\% of the
value of $h$ used in \cite{hqet:pap2}.}.

For all values of $\tilde{L}_1/a$ three hopping parameters $\kappa_{\rm h}$ have
then been fixed in order to achieve
\begin{equation}
z=\tilde{L}_1\,M\;=\;10.4, \; 12.1, \; 13.3 \,.
\end{equation}

\TABLEP{
\begin{tabular}{cccccc}
\hline\hline \\[-1.75ex]
$L/a$ & $z$ & \multicolumn{3}{c}{$R_1$} & $\Phi_2$  \\[0.5ex]
\hline\hline
     &        & $\theta_1=0$   &  $\theta_1=1/2$ &  $\theta_1=1$   & $\theta_0=0$\\
     &        & $\theta_2=1/2$ &  $\theta_2=1$   &  $\theta_2=0$   & \\
\hline
$20$ & $10.4$ &  $0.09795(13)$ &  $0.27426(30)$  &  $-0.37221(42)$ &    $7.847(40)$  \\
$20$ & $12.1$ &  $0.09512(12)$ &  $0.26588(30)$  &  $-0.36100(43)$ &    $9.108(46)$  \\
$20$ & $13.3$ &  $0.09336(12)$ &  $0.26068(30)$  &  $-0.35404(43)$ &    $10.068(50)$ \\
$24$ & $10.4$ &  $0.09958(18)$ &  $0.27904(37)$  &  $-0.37862(52)$ &    $7.697(44)$  \\
$24$ & $12.1$ &  $0.09689(17)$ &  $0.27110(37)$  &  $-0.36799(52)$ &    $8.866(50)$  \\
$24$ & $13.3$ &  $0.09528(17)$ &  $0.26632(36)$  &  $-0.36159(50)$ &    $9.716(54)$  \\
$32$ & $10.4$ &  $0.10157(30)$ &  $0.28481(71)$  &  $-0.38638(93)$ &    $7.512(53)$  \\
$32$ & $12.1$ &  $0.09897(30)$ &  $0.27717(71)$  &  $-0.37614(92)$ &    $8.623(58)$  \\
$32$ & $13.3$ &  $0.09744(30)$ &  $0.27265(71)$  &  $-0.37008(92)$ &    $9.411(62)$  \\
$40$ & $10.4$ &  $0.10283(30)$ &  $0.28806(52)$  &  $-0.39089(76)$ &    $7.484(51)$  \\
$40$ & $12.1$ &  $0.10027(30)$ &  $0.28052(52)$  &  $-0.38079(75)$ &    $8.575(56)$  \\
$40$ & $13.3$ &  $0.09876(29)$ &  $0.27608(52)$  &  $-0.37484(74)$ &    $9.344(60)$  \\
$CL$ & $10.4$ &  $0.10450(44)$ &  $0.29297(89)$  & $-0.39748(125)$ &    $7.341(96)$  \\ 
$CL$ & $12.1$ &  $0.10202(44)$ &  $0.28567(90)$  & $-0.38769(124)$ &    $8.386(102)$  \\
$CL$ & $13.3$ &  $0.10058(44)$ &  $0.28143(91)$  & $-0.38202(124)$ &    $9.106(107)$  \\

\hline\hline
\end{tabular}
\caption{Simulation results of the finite volume ($L=\tilde L_1$) relativistic observables 
needed in our main strategy. The continuum limits, obtained by linear extrapolation
in $(a/L)^2$ of the results for $L/a\geq 24$, are indicated by {\it CL}.
}\label{t:phi12}

}
\TABLEP{
\begin{tabular}{cccccccc}
\hline\hline \\[-1.75ex]
$L/a$ & $z$ & \multicolumn{3}{c}{$R_{\rm av}$} &  \multicolumn{3}{c}{$\tilde{\Phi}_3$}  \\[0.5ex]
\hline\hline
     &        & $\theta_1=0$   &  $\theta_1=1/2$ &  $\theta_1=1$   & $\theta_1=0$   &  $\theta_1=1/2$ &  $\theta_1=1$ \\
     &        & $\theta_2=1/2$ &  $\theta_2=1$   &  $\theta_2=0$   &                &                 & \\
\hline
$20$ & $10.4$ &   $0.1699(9) $ &  $0.4299(12)$ &  $-0.5998(20)$  &  $8.059(37)$  &  $8.293(37)$   & $ 8.993(37)$  \\
$20$ & $12.1$ &   $0.1668(9) $ &  $0.4198(11)$ &  $-0.5867(20)$  &  $9.315(37)$  &  $9.545(37)$   & $10.234(37)$  \\
$20$ & $13.3$ &   $0.1649(9) $ &  $0.4137(11)$ &  $-0.5787(19)$  &  $10.271(37)$ &  $10.500(37)$  & $11.180(37)$  \\
$24$ & $10.4$ &   $0.1739(23)$ &  $0.4391(31)$ &  $-0.6130(54)$  &  $7.864(39)$ &  $8.102(38)$  & $ 8.822(39)$ \\
$24$ & $12.1$ &   $0.1710(23)$ &  $0.4295(30)$ &  $-0.6005(52)$  &  $9.027(39)$ &  $9.263(38)$  & $ 9.971(39)$ \\
$24$ & $13.3$ &   $0.1693(22)$ &  $0.4239(29)$ &  $-0.5931(51)$  &  $9.874(39)$ &  $10.109(38)$ & $10.809(38)$ \\
$32$ & $10.4$ &   $0.1760(41)$ &  $0.4494(48)$ &  $-0.6254(90)$  &  $7.713(43)$ &  $7.941(41)$  & $8.661(42) $ \\
$32$ & $12.1$ &   $0.1733(40)$ &  $0.4403(46)$ &  $-0.6135(87)$  &  $8.818(42)$ &  $9.045(41)$  & $ 9.753(42)$ \\
$32$ & $13.3$ &   $0.1717(40)$ &  $0.4349(45)$ &  $-0.6066(85)$  &  $9.603(42)$ &  $9.828(41)$  & $10.531(42)$ \\
$40$ & $10.4$ &   $0.1790(70)$ &  $0.4493(72)$ &  $-0.6283(142)$ &  $7.656(45)$ &  $7.894(42)$  & $8.624(44) $ \\
$40$ & $12.1$ &   $0.1763(68)$ &  $0.4403(70)$ &  $-0.6166(138)$ &  $8.743(45)$ &  $8.979(42)$  & $9.698(44) $ \\
$40$ & $13.3$ &   $0.1747(67)$ &  $0.4352(68)$ &  $-0.6099(136)$ &  $9.509(45)$ &  $9.744(42)$  & $10.456(44)$ \\
$CL$ & $10.4$ &   $0.1801(75)$ &  $0.4587(84)$ &  $-0.6392(159)$ &  $7.533(89)$ &  $7.765(86)$  & $8.496(88)$ \\
$CL$ & $12.1$ &   $0.1776(73)$ &  $0.4502(81)$ &  $-0.6280(154)$ &  $8.573(91)$ &  $8.805(88)$  & $9.524(89)$ \\
$CL$ & $13.3$ &   $0.1761(72)$ &  $0.4452(79)$ &  $-0.6218(151)$ &  $9.289(93)$ &  $9.519(91)$  & $10.234(92)$\\ 
\hline\hline
\end{tabular}
\caption{
Same as Table \ref{t:phi12} in the case of the alternative strategy.
}\label{t:phi23tilde}

}

We collect these parameters in \tab{t:qcdpar}, whereas the results
for the quantities needed in the matching step are summarized in
Tables~\ref{t:phi12} and~\ref{t:phi23tilde}. The  errors there include
systematics due to the uncertainties in the $Z$-factors, in
particular, the error on the universal factor $h$ has been propagated
only {\em after} performing the continuum limit extrapolations.
Ensembles of roughly 2000 (for $\tilde{L}_1/a=20$) to few hundreds (for $\tilde{L}_1/a=40$)  
gauge configurations have been generated for this part of
the computation.
The lattice $\tilde{L}_1/a=20$ is not used in the extrapolations but rather to check
for the smallness of higher order cutoff effects for $\tilde{L}_1/a \geq 24$.

Concerning the simulation of HQET, we have computed the various quantities 
in the two required volumes. 
The first one, where we match the effective theory with QCD, 
has a space extent $L_1$. 
The second one is such that $L_2=2 L_1$.
The value of the \SF renormalized coupling is fixed at $\bar{g}^2(L_1) = 3.48$, 
and we have used the
resolutions $L_1/a = 6, 8, 10, 12$. The corresponding values of 
$\beta$ as well as $\kappa=\kappa_c$ 
can be found in Table A.1 of  \cite{hqet:pap1}.
All these quantities are computed with two different actions, 
HYP1 and HYP2. The continuum values are then obtained by 
constraining the fits to give the same values for these actions.
We note that the results for HYP1 and HYP2 are statistically correlated.

\TABLET{
\begin{tabular}{ccccccccc}
\hline\hline \\[-1.75ex]
$L/a$ & \multicolumn{8}{c}{$R_1^{\rm stat}$} \\[0.5ex]
\hline\hline
& \multicolumn{2}{c}{$\theta_1=0$}   &&  \multicolumn{2}{c}{$\theta_1=1/2$} && \multicolumn{2}{c}{$\theta_1=1$} \\
& \multicolumn{2}{c}{$\theta_2=1/2$} &&  \multicolumn{2}{c}{$\theta_2=1$} &&  \multicolumn{2}{c}{$\theta_2=0$} \\
& \hypone & \hyptwo &&  \hypone & \hyptwo &&  \hypone & \hyptwo \\[0.5ex]
\hline
6   & 0.06936(5) & 0.06939(4) && 0.18583(7)  &  0.18591(7)  && -0.25519(12) & -0.25530(11)  \\
8   & 0.07572(6) & 0.07574(6) && 0.20452(11) &  0.20457(11) && -0.28024(17) & -0.28031(17)  \\
10  & 0.07821(5) & 0.07822(5) && 0.21246(8)  &  0.21249(8)  && -0.29067(13) & -0.29071(13) \\
12  & 0.07934(8) & 0.07935(8) && 0.21622(13) &  0.21625(13) && -0.29556(21) & -0.29559(20) \\
{\it CL} & \multicolumn{2}{c}{ 0.08238(12) } && \multicolumn{2}{c}{ 0.22596(21) } &&  
\multicolumn{2}{c}{  -0.30835(32) }\\
\hline\hline
\end{tabular}
\caption{
Lattice results of $R_1^{\rm stat}$ for $L=L_1$.
The continuum limits are obtained by a linear extrapolation
in $(a/L)^2$ of the results for $L/a\geq 8$.
}\label{t:R1stat}

}

\TABLET{
\begin{tabular}{ccccccccc}
\hline\hline \\[-1.75ex]
$L/a$ & \multicolumn{8}{c}{$R_{\rm av}^{\rm stat}$} \\[0.5ex]
\hline\hline
& \multicolumn{2}{c}{$\theta_1=0$}   &&  \multicolumn{2}{c}{$\theta_1=1/2$} && \multicolumn{2}{c}{$\theta_1=1$} \\
& \multicolumn{2}{c}{$\theta_2=1/2$} &&  \multicolumn{2}{c}{$\theta_2=1$} &&  \multicolumn{2}{c}{$\theta_2=0$} \\
& \hypone & \hyptwo &&  \hypone & \hyptwo &&  \hypone & \hyptwo \\[0.5ex]
\hline
%
6   &  0.1502(3) & 0.1543(3) && 0.3562(3) &  0.3688(3) && -0.5231(6)   & -0.5231(6) \\ 
8   &  0.1544(4) & 0.1575(4) && 0.3672(4) &  0.3765(4) && -0.5216(7)   & -0.5340(8) \\
10  &  0.1571(5) & 0.1595(5) && 0.3724(6) &  0.3710(6) && -0.5295(10)  & -0.5391(10) \\
12  &  0.1561(8) & 0.1579(8) && 0.3729(8) &  0.3786(9) && -0.5289(15)  & -0.5365(16) \\
{\it CL} & \multicolumn{2}{c}{ 0.1606(6) } && \multicolumn{2}{c}{ 0.3827(6)  } &&  
\multicolumn{2}{c}{  -0.5432(11) }\\
\hline\hline
\end{tabular}
\caption{
Lattice results of $R_{\rm av}^{\rm stat}$.
The details are the same as in Table \ref{t:R1stat}.
}\label{t:Ravstat}

}

For the computation of the step scaling functions one uses
the same $\beta,\kappa$ and $L_2/a=2L_1/a$.
All these computations are done with several thousand gauge configurations.
Note that, even if $L_1$ is the same in QCD and in HQET, 
the typical lattice spacings are much larger in the effective
theory.
The results of $R_1^{\rm stat}$ and $R_{\rm av}$ can be found in 
Tables~\ref{t:R1stat} and~\ref{t:Ravstat}. The values of the
step scaling functions are collected in Tables~\ref{t:Sigmam},
\ref{t:Sigmakin1} and~\ref{t:Sigmakin2}.

\TABLET{
\begin{tabular}{cccccc}
\hline\hline \\[-1.75ex]
$L/a$ & \multicolumn{2}{c}{$\Sigmam(3.48,a/L)$}\\[0.5ex]
\hline\hline
& \hypone &   \hyptwo \\[0.5ex]
\hline
 8  & 0.431(11) & 0.411(11) \\
 10 & 0.437(11) & 0.424(10) \\
 12 & 0.422(16) & 0.418(16) \\
{\it CL} & \multicolumn{2}{c}{0.430(25)} \\

\hline\hline
\end{tabular}
\caption{
Lattice results of the step scaling function $\Sigmam$.
The bare parameters are described in the text. 
The continuum limit is obtained by a linear extrapolation
in $(a/L)^2$ of the results for $L/a\geq 8$.
}\label{t:Sigmam}

}
\TABLET{
\begin{tabular}{ccccccccc}
\hline\hline \\[-1.75ex]
$L/a$ & \multicolumn{8}{c}{$\Sigma_1^\mrm{kin}(3.48,a/L)$} \\[0.5ex]
\hline\hline
& \multicolumn{2}{c}{$\theta_1=0$}   &&  \multicolumn{2}{c}{$\theta_1=1/2$} && \multicolumn{2}{c}{$\theta_1=1$} \\
& \multicolumn{2}{c}{$\theta_2=1/2$} &&  \multicolumn{2}{c}{$\theta_2=1$} &&  \multicolumn{2}{c}{$\theta_2=0$} \\
& \hypone & \hyptwo &&  \hypone & \hyptwo &&  \hypone & \hyptwo \\[0.5ex]
\hline
 6 & 0.6241(17) & 0.6245(11)  && 0.6219(60) & 0.6223(5)  && 0.6225(8)  & 0.6228(6)  \\ 
 8 & 0.5790(20) & 0.5797(13)  && 0.5789(65) & 0.5793(5)  && 0.5789(10) & 0.5794(7)  \\
10 & 0.5587(47) & 0.5586(22)  && 0.5585(14) & 0.5588(9)  && 0.5586(22) & 0.5590(14) \\
12 & 0.5364(66) & 0.5342(39)  && 0.5424(19) & 0.5417(12) && 0.5409(30) & 0.5398(18) \\
{\it CL} & \multicolumn{2}{c}{0.457(10)} && \multicolumn{2}{c}{0.471(3)} &&  \multicolumn{2}{c}{0.467(5)}\\
\hline\hline
\end{tabular}
\caption{
Lattice results of the step scaling function $\Sigma_1^\mrm{kin}$.
The continuum limits are obtained by a linear extrapolation
in $a/L$ of the results for $L/a\geq 8$.
}\label{t:Sigmakin1}

}
\TABLET{
\begin{tabular}{ccccccccc}
\hline\hline \\[-1.75ex]
$L/a$ & \multicolumn{8}{c}{$\Sigma_2^\mrm{kin}(3.48,a/L)$} \\[0.5ex]
\hline\hline
& \multicolumn{2}{c}{$\theta_1=0$}   &&  \multicolumn{2}{c}{$\theta_1=1/2$} && \multicolumn{2}{c}{$\theta_1=1$} \\
& \multicolumn{2}{c}{$\theta_2=1/2$} &&  \multicolumn{2}{c}{$\theta_2=1$} &&  \multicolumn{2}{c}{$\theta_2=0$} \\
& \hypone & \hyptwo &&  \hypone & \hyptwo &&  \hypone & \hyptwo \\[0.5ex]
\hline
8  & 4.81(44) & 4.72(32) &&  1.58(15) & 1.55(10) && -1.19(11) & -1.17(8)  \\
10 & 4.34(58) & 4.20(39) &&  1.43(19) & 1.39(13) && -1.08(15) & -1.04(10) \\
12 & 4.79(86) & 3.98(58) &&  1.58(28) & 1.31(19) && -1.19(21) & -0.99(14) \\

{\it CL} & \multicolumn{2}{c}{2.9(1.5)} && \multicolumn{2}{c}{0.96(50)} &&  \multicolumn{2}{c}{-0.71(38)}\\
\hline\hline
\end{tabular}
\caption{
Same as Table \ref{t:Sigmakin1} for $\Sigma_2^\mrm{kin}$.
}\label{t:Sigmakin2}

}
Finally there are simulations in small volume to obtain the subtractions
$\meffonestat(L_2)$ and $\meffonekin(L_2)$. These are done with $L_2=1.436\,r_0$
and $\beta$ determined from the knowledge of $r_0/a$ \cite{pot:intermed}.
The parameters, including $\kappa=\kappa_c$, are listed in 
Table 6 of \cite{stat:actpaper}. The values of $\beta$ do of course agree 
with the ones employed in the large volume, which we describe in the 
next appendix.



\section{Large volume simulations and extraction 
        of matrix elements \label{s:large}}

The parameters for the simulations in large volume are collected in
Table~\ref{t:infvol} together with the results for $E_{\rm stat}$ and
$E_{\rm kin}$. The lattice extension $L/a$ and $\beta$ are such that
$L=4L_1 \approx 3/2\,\fm$ except for the second lattice where 
we have $L=6L_1 \approx 2\,\fm$. This lattice is used only to check
for the absence of finite size effects. 
We see from Table~\ref{t:infvol} that finite size effects are indeed very small,
the difference between the results from the $L/a=16$ and the $L/a=24$
lattices at $\beta=6.0219$ is consistent with zero within at most one
standard deviation ($aE_{\rm stat}$ from HYP1).
The number of
configurations generated ranges from $4300$ at $\beta=6.0219$ to $2200$ at
$\beta=6.4956$ (for the larger volume at $\beta=6.0219$ we had $1300$ configurations).
Since our phenomenological input is 
the mass of the (spin averaged) $B_{\rm s}$
meson, we set $\kappa$ to $\kappa_{\rm s}$
in order to reproduce the quenched value of the strange quark mass
from Ref.~\cite{mbar:pap3}, i.e.
\begin{equation}
M_{\rm s}r_0=0.35(1)\;, 
\end{equation}
with $M_{\rm s}$ the renormalization group invariant strange quark
mass defined as in Appendix~\ref{s:sim} after replacing $\kappa_{\rm
  h}$ by $\kappa_{\rm s}$.

\TABLE{
\begin{tabular}{ccccccc}
\hline\hline \\[-1.75ex]
$\beta$   &   $\kappa_{\rm s}$  &  $L^3\times T$           &  \multicolumn{2}{c}{$aE_{\rm stat}$}   & \multicolumn{2}{c}{$a
^2E_{\rm kin}$}  \\[0.5ex]
\hline\hline
          &                     &                          &  HYP1 &
          HYP2 & HYP1 & HYP2 \\[0.5ex]
\hline
$6.0219$  &   $0.133849$        & $16^3 \times 24\; [32]$  & $0.4345(21)$
  & $0.4029(32)$   & $0.750(4)$   &  $0.774(3)$ \\
$6.0219$  &   $0.133849$        & $24^3 \times 32$  & $0.4378(25)$
  & $0.4034(20)$   & $0.746(7)$   &  $0.776(5)$ \\
$6.2885$  &   $0.1349798$       & $24^3 \times 48$ & $0.3295(21)$
 & $0.3034(29)$ &$0.643(7)$   & $0.676(5)$   \\
$6.4956$  &   $0.1350299$       & $32^3\times 64$ & $0.2724(20)$
& $0.2461(14)$ & $0.599(10)$  &$0.620(11)$   \\
\hline\hline
\end{tabular}
\caption{Parameters of the large volume  simulations. Where present, the numbers in
  brackets refer to a second dataset at the same $(\beta,\kappa)$ values.
}\label{t:infvol}

}

The numbers for $E_{\rm stat}$ and $E_{\rm kin}$ have been obtained by
applying two different fitting procedures to two independent datasets
(where available). The quoted errors are such that both the results
are covered and they therefore provide a reasonable estimate
of the systematics associated with the fits. We now sketch these
procedures.

Let us consider in QCD the effective ``mass'' $\Gamma(x_0)$ obtained from the
correlation function $F_{\rm av}(x_0)$ in eq.~(\ref{e:fav}) and its quantum-mechanical decomposition 
\be
\Gamma(x_0)=-{{\partial_0+\partial^*_0}\over{2}} F_{\rm av}=E_0+A
e^{-\Delta x_0}+ \dots
\ee
where $E_0$ is the energy of the ground state, $\Delta$ is the gap
between the ground and the first excited states and  the dots refer to contributions 
from higher states. The $1/m_{\rm b}$ expansion reads
\begin{eqnarray}
\label{gammaexp}
\Gamma(x_0)&=& E_{\rm stat}+ \omega_{\rm kin}E_{\rm kin}+ (A^{\rm
  stat}+\omega_{\rm kin}A^{\rm kin})e^{-\Delta^{\rm
    stat}x_0}(1-\omega_{\rm kin}x_0\Delta^{\rm kin}) + \dots \nonumber \\ 
          &=& \Gamma^{\rm stat}(x_0)+\omega_{\rm kin}\Gamma^{\rm
            kin}(x_0) +\dots 
\end{eqnarray}
where $ \Gamma^{\rm stat}$ and $ \Gamma^{\rm kin}$ are defined in
analogy to eqs.~(\ref{e:meffone}, \ref{e:meffonekin}) 
in terms of the correlators $f_{\rm A}^{\rm
  stat}(x_0)$ and $f_{\rm A}^{\rm kin}(x_0)$. 

In the correlation
function $f_{\delta{\rm A}}(x_0)$ the same states contribute as in
$f_{\rm A}(x_0)$.  Performing again first the quantum-mechanical
decomposition and then the $1/m_{\rm b}$ expansion of these
correlators, it is easy to see that the ratios
\be
P_{\rm A}^{\rm stat}(x_0)= {{f_{\rm A}^{\rm stat}(x_0)}\over{f_{\delta
      \rm A}^{\rm stat}(x_0)}} \quad {\rm and} \quad P_{\rm A}^{\rm
  kin}(x_0)=P_{\rm A}^{\rm stat}(x_0)\left[{{f_{\rm A}^{\rm
        kin}(x_0)}\over{f_{\rm A}^{\rm stat}(x_0)}}-
{{f_{\delta \rm A}^{\rm kin}(x_0)}\over{f_{\delta \rm A}^{\rm
      stat}(x_0)}} \right] 
\label{e:R}
\ee
have the following form
\bea
\label{pstatexp}
P_{\rm A}^{\rm stat} &=& b_1+b_2 e^{-\Delta^{\rm stat}x_0} \;, \\
\label{pkinexp}
P_{\rm A}^{\rm kin}  &=& b_3+b_4 e^{-\Delta^{\rm stat}x_0}
-b_2\Delta^{\rm kin}x_0 e^{-\Delta^{\rm stat}x_0} \;. 
\eea
They 
can therefore be used to further constrain 
$\Delta^{\rm stat}$ and $\Delta^{\rm kin}$.
We are thus lead to perform a combined fit
\bea
\Gamma^{\rm stat} &=& b_5+b_6 e^{-\Delta^{\rm stat}x_0} \;, \\
\Gamma^{\rm kin} &=& b_7 +b_8 e^{-\Delta^{\rm stat}x_0} -b_6
\Delta^{\rm kin} x_0 e^{-\Delta^{\rm stat}x_0} \;,
\eea
together with eq. (\ref{pstatexp}) and (\ref{pkinexp}), 
with non-linear parameters $a_1=\Delta^{\rm stat}$ and 
$a_2=\Delta^{\rm kin}$ and the linear
parameters $b_i$, which contain the desired
$b_5=E_{\rm stat}$ and $b_7=E_{\rm kin}$. 

Since the correction terms 
are nevertheless not so easy to compute 
at the smaller lattice spacings, we perform the above fit first
at $\beta=6.0219$ and extract $a\Delta^{\rm stat}$ and $a^2\Delta^{\rm kin}$.
We then use that these quantities scale {\it roughly} 
(i.e. $r_0\,\Delta^{\rm stat}\approx$constant and 
$r_0^2\,\Delta^{\rm kin}\approx$constant). 
To implement this,
we input the scaled means as priors~\cite{Lepage:priors} in a
second step where we add
\be
\chi^2_{\rm prior} =\sum_{i=1,2} {{\left(a_i-a_1^{\rm
        prior}\right)^2}\over {(\delta a_i^{\rm prior})^2}} \;, 
\ee
to the standard $\chi^2$.  The uncertainty 
 $\delta a_i^{\rm prior}$ is taken from the 
fit result at $\beta=6.0219$. However, in order to remain on the safe side, 
it  is not scaled but kept
constant at the smaller
lattice spacing. Thus  $\delta a_2^{\rm prior}/a_2^{\rm prior} \propto 1/a^2$ 
for example. The constraint due to the priors becomes weaker as we
approach the continuum. 

Here and in the following procedure the fit range is chosen to keep
a minimum physical distance from the boundaries, namely 
$x_0 \geq t_{\rm  min} \approx 2r_0$. The stability of the results is checked by varying
$t_{\rm min}$ to $t_{\rm min} -r_0/2$.
As an example we show in figure~\ref{f:fit} the
results for $P_{\rm A}^{\rm stat},\,P_{\rm A}^{\rm kin},\,\meffstat$ and $\Gamma^{\rm kin}$ at
$\beta=6.2885$. 
One observes that $P_{\rm A}^{\rm stat},\, \meffstat$ provide
very good constraints of the parameters $\Delta^{\rm stat},\,b_2,\,b_6$. The
remaining ones are then effectively linear fit
parameters. Nevertheless, the 
error band of $\Ekin$ (dashed line) resulting from the fit  
is not that small. 

%
\FIGURE{
\includegraphics*[width=6.9cm]{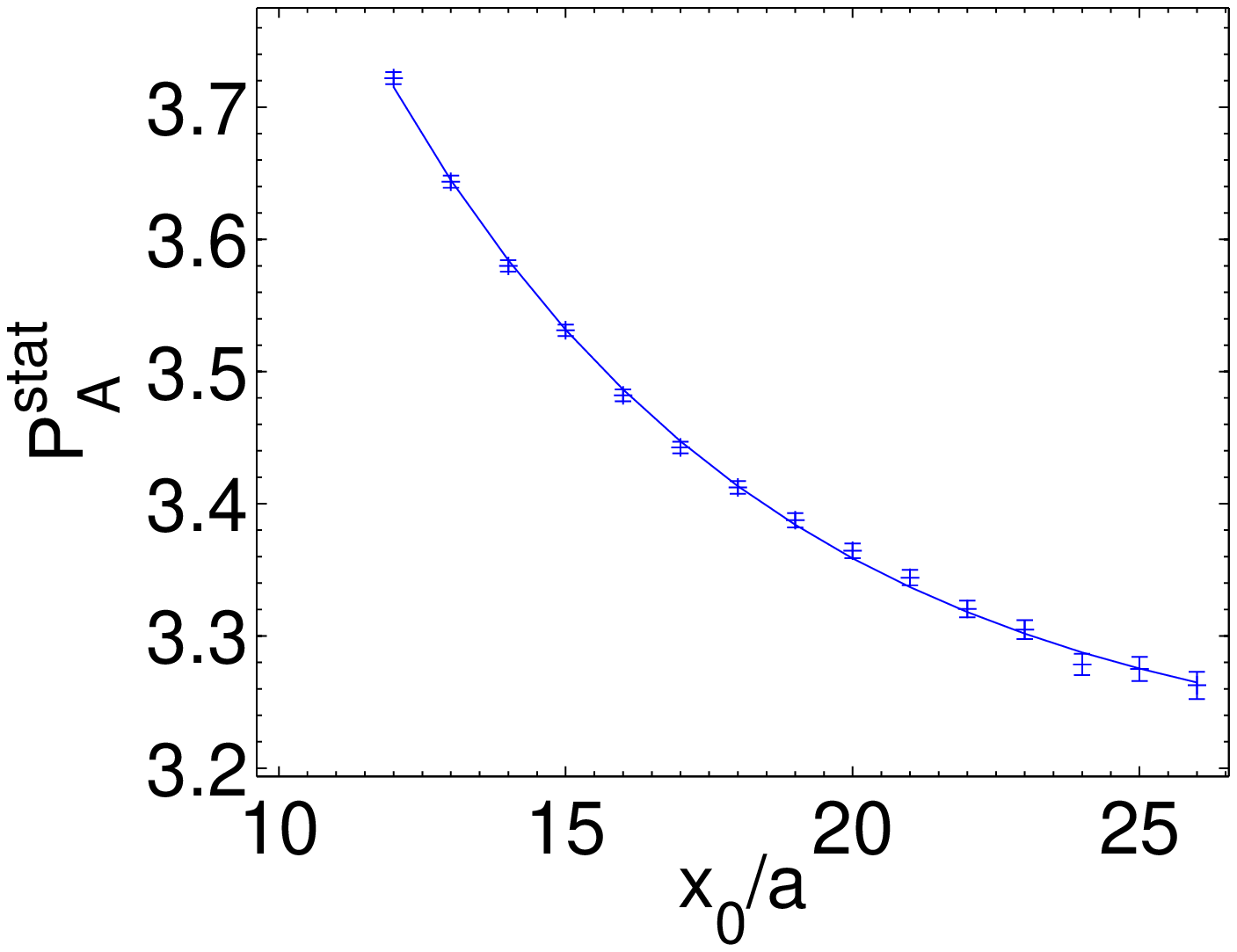}
\hspace{0.5cm}
\includegraphics*[width=6.9cm]{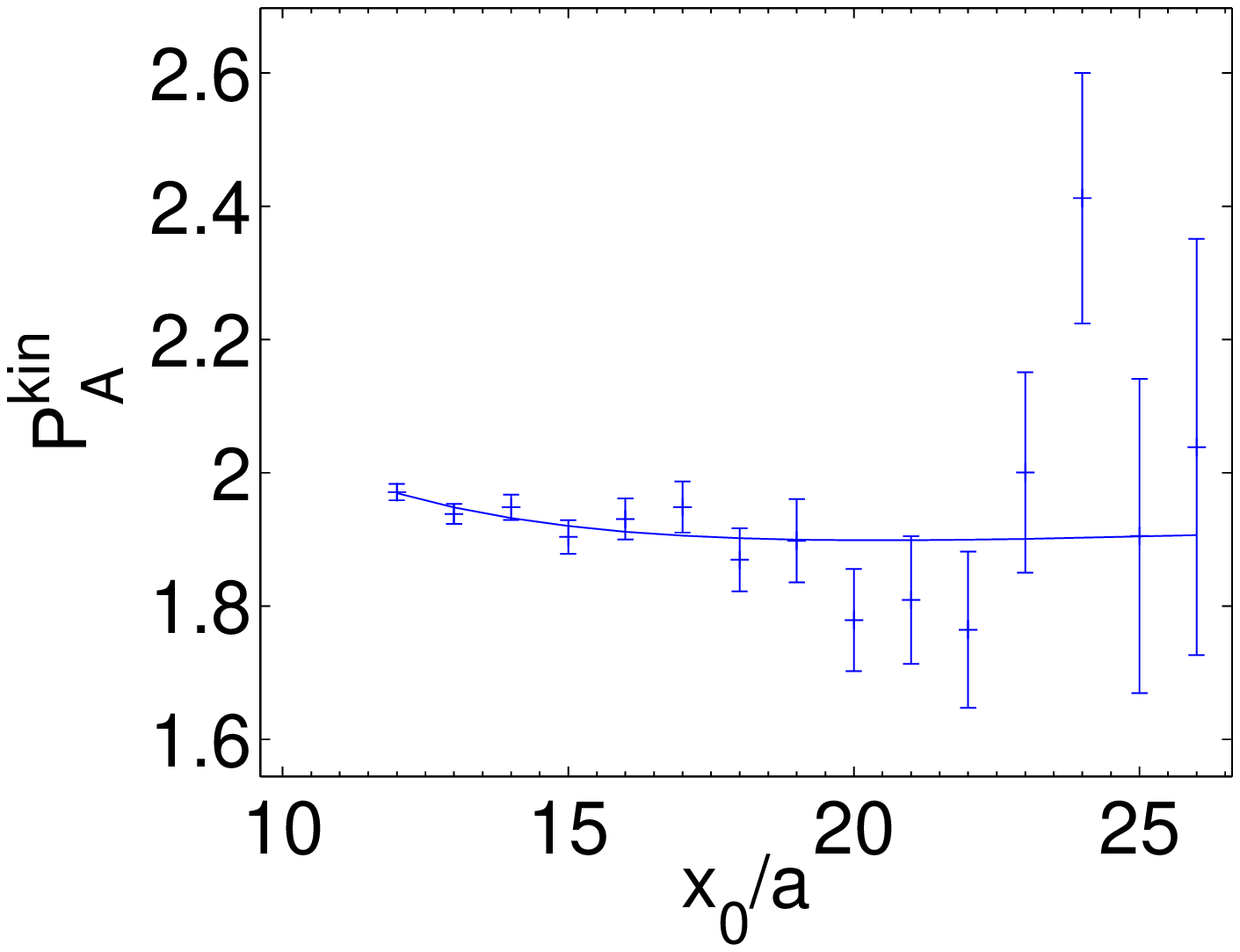} 
\includegraphics*[width=6.9cm]{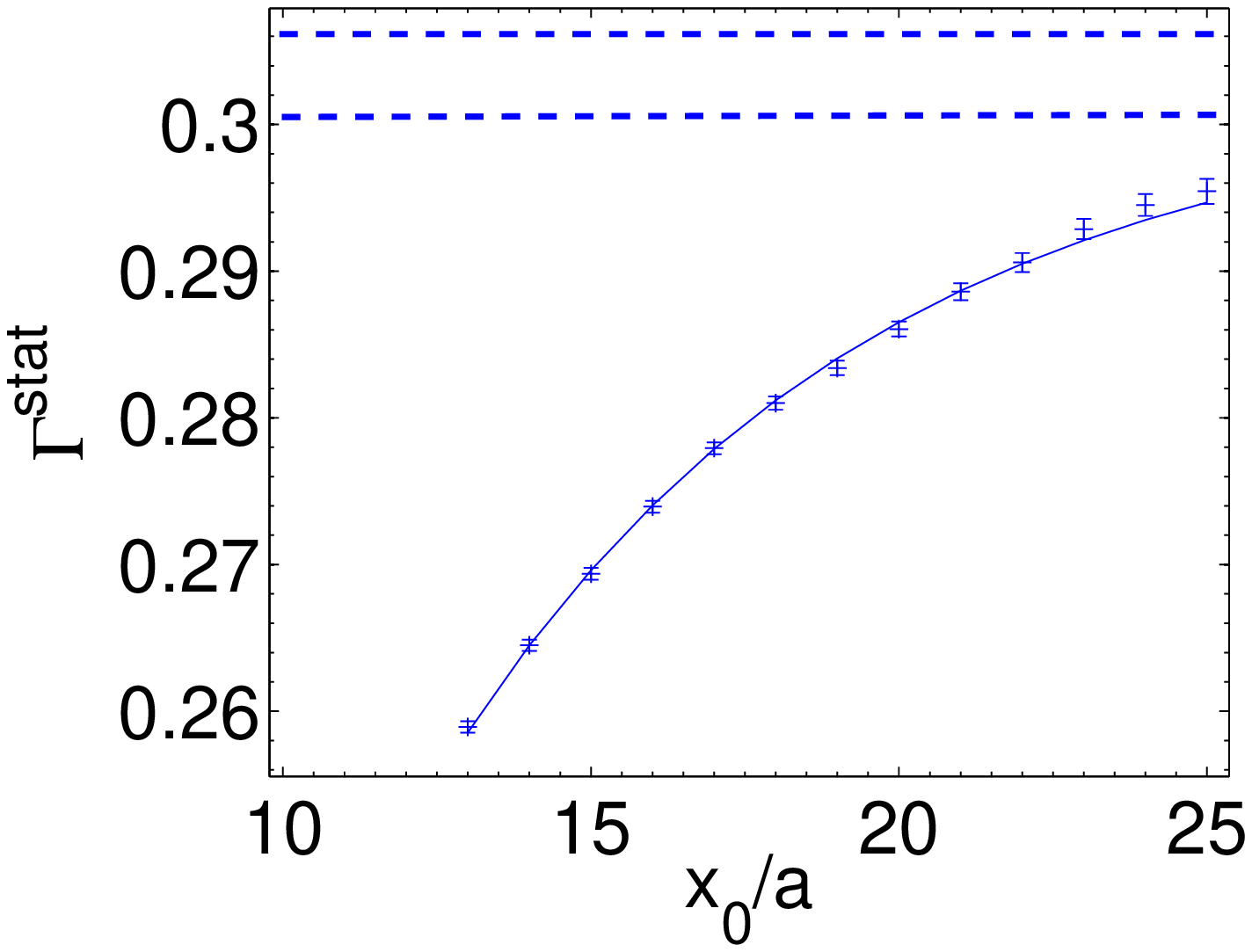}
\hspace{0.5cm}
\includegraphics*[width=6.9cm]{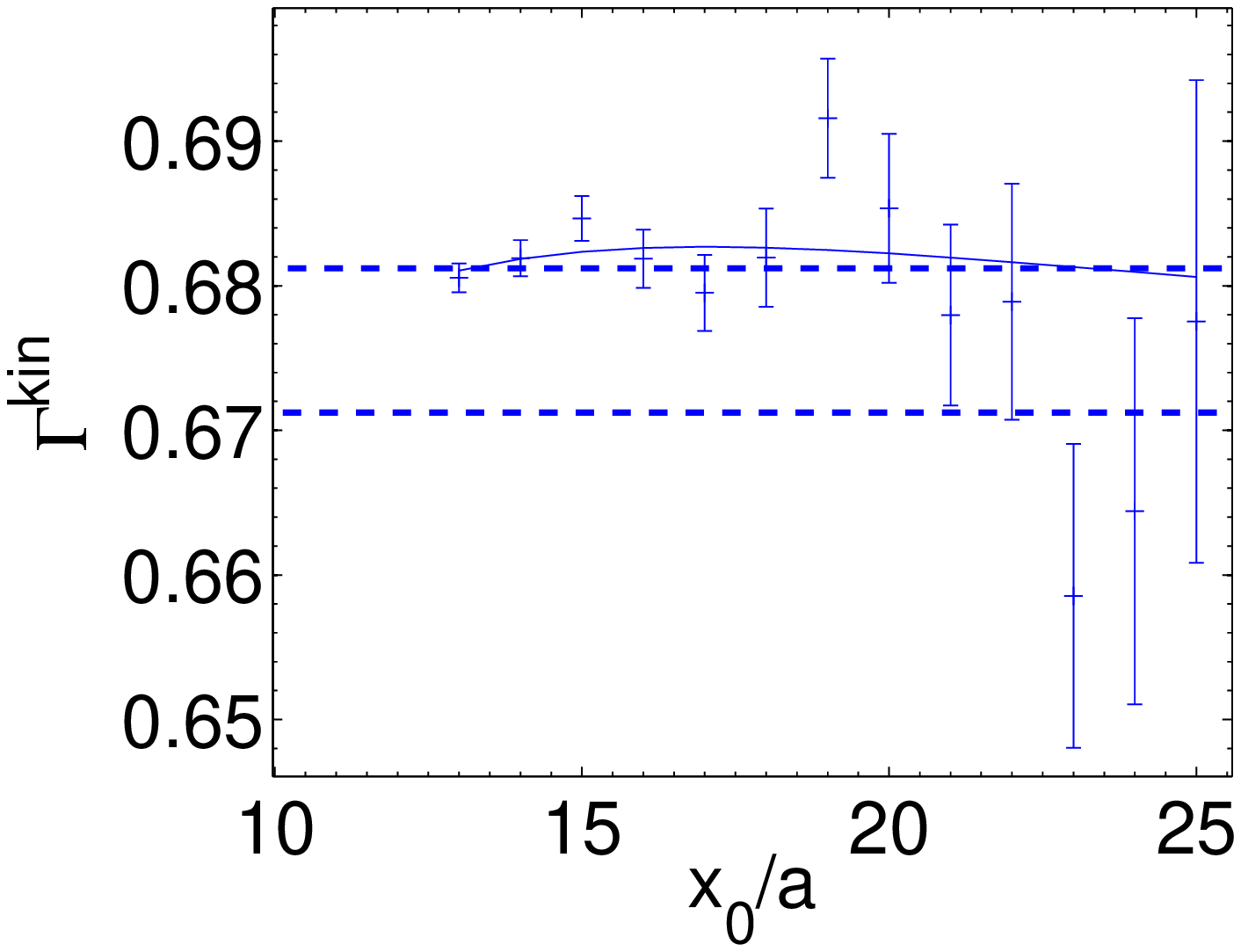}
\caption{Results for $P_{\rm A}^{\rm stat}$, $P_{\rm A}^{\rm kin}$, 
$\Gamma^{\rm stat}$ and $\Gamma^{\rm kin}$ at
$\beta=6.2885$ (HYP2) with the corresponding functions obtained by the fit.
}\label{f:fit}
}

An alternative strategy is used to get a second estimate of $E_{\rm
  stat}$ at the two coarser lattice spacings, where we have two
independent datasets. Exploiting again the remark before eq.~(\ref{e:R})
we construct an effective mass $\Gamma_{\delta \rm A}$ from the correlator $f_{\delta \rm
  A}(x_0)$ in the very same way as $\Gamma^{\rm stat}$ is obtained from
$f_{\rm A}^{\rm stat}(x_0)$. The idea is to combine the two effective
masses in order to eliminate the contribution from the first excited
state and then perform a fit to a constant (in the mentioned fit
range). In practice we minimize the quantity
\be
Q=\sum_{x_0=t_{\rm min}}^{t_{\rm max}} \left[\Gamma^{\rm stat}(x_0)+
  \alpha \Gamma_{\delta {\rm A}}(x_0) -B\right]^2 \;, 
\ee
with respect to $\alpha,\;B$. Finally the weighted average of $\left[\Gamma^{\rm stat}(x_0)+
  \alpha \Gamma_{\delta {\rm A}}(x_0)\right] /(1+\alpha)$ yields the
estimate of $E^{\rm stat}$. The quality of the result is comparable to
that obtained in the first approach.


\section{Alternative strategy  \label{s:alt}}

We briefly introduce an alternative strategy, based 
on the correlation functions $\fa,\kv$ in addition to
$\fone,\kone$. With
\bes
 F_\mrm{av}(x_0,\theta) = {1\over4}\log \left( - \left[\fa\right]_\mrm{R}(x_0,\theta)
                           \times \left[\kv\right]_\mrm{R}^3(x_0,\theta)\right)
\label{e:fav}
\ees
we introduce 
\bes
  R_\mrm{av}(L,\theta_1,\theta_2) &=& F_\mrm{av}(x_0,\theta_1)-F_\mrm{av}(x_0,\theta_2) \atxhalf \\
  \meffav(L,\theta_0) &=& -{\partial_0 +  \partial_0^* \over 2} F_\mrm{av}(x_0,\theta_0) \atxhalf\,.
\ees
Keeping $\Phi_1$, from the standard strategy, we define the set of observables
\bes
  \widetilde\Phi_1 (L,\Mbeauty) &=& \Phi_1 (L,\Mbeauty)\,,\\
  \widetilde\Phi_2 (L,\Mbeauty) &=& R_\mrm{av}(L,\theta_1,\theta_2) - R_\mrm{av}^\mrm{stat}(L,\theta_1,\theta_2)
        \,,\\
  \widetilde\Phi_3 (L,\Mbeauty) &=& L \meffav(L,\theta_0)\,,
\ees
with the $\minv$ expansion
\bes
  \widetilde\Phi_2 (L,\Mbeauty) &=& \omegakin \ratakin(L,\theta_1,\theta_2) 
        +\cavhqet \ratdastat(L,\theta_1,\theta_2)\,\\
  \widetilde\Phi_3(L,\Mbeauty) &=& L\,\big[ \mhbare + \meffstat(L,\theta_0) 
                               + \omegakin\meffkin(L,\theta_0) + \cavhqet \meffdastat(L,\theta_0)\big]\,, 
\ees
where due to the spin average the combination
\bes
  \cavhqet = {1\over4}[\cahqet+3\cvhqet]
\ees
is present. The so far undefined terms $R_\mrm{av}^\mrm{stat}, \ratakin,\meffkin,\ratdastat,\meffdastat$ are
straightforwardly obtained from our definitions. 

%
The alternative observables change from $L$ to $2L$ via
\bes
 \widetilde\Phi_i(2L,\Mbeauty) &=&
        \sum_{j\leq i} \sigma_{ij}(u) \,\widetilde\Phi_j(L,\Mbeauty)
        \;+\; \delta_{i3} \; \widetilde\sigmam(u)\,, \\
  \sigma_{ij}(u) &=& \lim_{a/L \to 0} \Sigma_{ij}(u,a/L) 
\ees
with the step scaling functions (we drop arguments $\theta_1,\theta_2$
and $u=\gbar^2(L)$ is understood)

\bes
  \Sigma_{11}(u,a/L) &=& \ratonekin(2L)/\ratonekin(L) =  \Sigma_{1}^\mrm{kin}(u,a/L)
                 \\[1ex]
  \Sigma_{21}(u,a/L) 
                &=&  {1 \over \ratonekin(L) }
                \{\ratakin(2L) - \ratakin(L) \,\Sigma_{22}(u,a/L) \}
                  \\
  \Sigma_{22}(u,a/L) &=& \ratdastat(2L)/\ratdastat(L)
                  \\[1ex]
  \Sigma_{31}(u,a/L) 
                &=&  {2L  \{\meffkin(2L) - \meffkin(L)\}
                        \over \ratonekin(L) } -
                 \, \Sigma_{32}(u,a/L)
                        {\ratakin(L) \over \ratonekin(L) }
                  \\
  \Sigma_{32}(u,a/L) &=&  2L
                {\meffdastat(2L)-\meffdastat(L) \over \ratdastat(L)}
                  \\
  \Sigma_{33}(u,a/L) &=&  2 \\[1ex]
   \widetilde\sigmam(u) &=&  \lim_{a/L \to 0} 2L\,\big[\meffstat(2L) - \meffstat(L) \big]\,.
\ees 
The final relation for the B-meson mass is  \eq{e:mbavsplit} with
\bes
  \label{e:masteralt}
   L_2\mB^\mrm{(0a)}(\Mbeauty) &=&  \widetilde\sigmam(u_1) + 2\,\widetilde\Phi_3(L_1,\Mbeauty) \,,\\
   L_2\mB^\mrm{(0b)}(\Mbeauty) &=& L_2 [E^\mrm{stat} - \meffstat(L_2)]\,, \\
 L_2\mb^{(1a)}(\Mbeauty) &=& \sigma_{31}(u_1)\,\widetilde\Phi_1(L_1,\Mbeauty)
                            +\sigma_{32}(u_1)\,\widetilde\Phi_2(L_1,\Mbeauty)\,,\\
 L_2\mb^{(1b)}(\Mbeauty) &=&  L_2\left[ {E^\mrm{kin} - \meffkin(L_2) \over \ratonekin(L_2)} +  
         {\meffdastat(L_2 )\, \ratakin(L_2) \over  \ratdastat(L_2)\, \ratonekin(L_2) } \right]
         \,\sigmakin_1(u_1)\,\widetilde\Phi_1(L_1,\Mbeauty) \nonumber \\
                         &&   - L_2 {\meffdastat(L_2 ) \over \ratdastat(L_2)}
\left[ \sigma_{21}(u_1)\,\widetilde\Phi_1(L_1,\Mbeauty) + \sigma_{22}(u_1)\,\widetilde\Phi_2(L_1,\Mbeauty) \right]\,.
\ees

Although the results have been already given in \tab{t:Mb_sum}, the 
reader will find more details in \tab{t:Mb_3b3}.
\TABLE{
\begin{tabular}{cccccccccccc}
\hline\hline \\[-1.75ex]
$\theta_0$ &  $r_0\Mb^{(0)}$ && 
\multicolumn{4}{c}{$r_0\Mb^{(1a)}$}  & \multicolumn{3}{c}{$r_0\Mb^{(1b)}$} & \\  
\hline\hline
& &&
$\theta_1=0$   &  $\theta_1=1/2$ & $\theta_1=1$ &&
$\theta_1=0$   &  $\theta_1=1/2$ & $\theta_1=1$ \\
& &&
$\theta_2=1/2$ &  $\theta_2=1$   & $\theta_2=0$ &&
$\theta_2=1/2$ &  $\theta_2=1$   & $\theta_2=0$ \\
\hline
0   & 17.05(25) && 0.17(6)  & 0.17(6)  & 0.17(6) && 0.02(9)  & 0.02(8) & 0.02(9)  \\
1/2 & 17.01(22) && 0.20(7)  & 0.18(6)  & 0.19(7) && 0.02(10) & 0.02(9) & 0.02(9)  \\
1   & 16.78(28) && 0.34(11) & 0.30(7)  & 0.32(8) && 0.06(12) & 0.06(9) & 0.06(10) \\
\hline\hline
\end{tabular}
\caption{
RGI results of $\Mb$ in the static approximation and of the $\minv$ correction
for the alternative strategy.
}\label{t:Mb_3b3}

}


\section{Propagating uncertainties in 
	$L_i/r_0$ and $\gbar^2(L_i)$ \label{s:shift}}

In our simulations we have fixed $\tilde L_1$ by 
$\gbar^2(\tilde L_1/4)=1.8811$, because the corresponding
bare parameters $\beta,\kappa$ are available in the literature.
We here give the estimate of the small effect caused by 
$\tilde L_1\neq L_1$ in the static approximation. From the polynomial 
interpolations of the step scaling function
of the coupling, $\sigma(u)$ \cite{mbar:pap1}, we estimate the corresponding
mismatch in couplings as
\bes
 \tilde u -u = \gbar^2(\tilde L_1) - \gbar^2(L_1) =  
  \sigma(\sigma(1.8811)) -3.48 = -0.17(5)\,.
\ees
Let us write 
\bes
  {\Mbeauty \over \mB} = \rho(\tilde u,z)  \, [1 + K(u)]  \quad \mbox{at} \quad \tilde u=u
\ees
with
\bes
  K(u)	= {\meffonestat(L_1) - \Estat \over \mB} \,, \quad
  \rho(u,z)=  {z \over \Phi_2(u,z)}\,.
\ees
The relation ${\rmd\over\rmd u} {\Mbeauty \over \mB} = 0$ gives
\bes
  -{1+K(u) \over \rho(u,z)} {\rmd\over\rmd u} \rho(u,z)  = K'(u) =
  {1 \over \mB} {\rmd\over\rmd u} \meffonestat \,.
\ees
Denoting by $\Delta\Mbeauty$ the correction we have to add to 
$\Mbeauty$ when it is computed with $\tilde u \neq u$ (as we did), 
we get from the above equations
\bes
 {1\over \mB} \Delta\Mbeauty =
 [\tilde u - u] \times  \rho(u) K'(u) \,,
\ees
where $K'(u)$ is easily estimated by taking a numerical derivative
of $\meffonestat$. From the difference of 
$L/a=12$ and $L/a=10$ at fixed $g_0^2$ (with $\gbar^2|_{L/a=12}=3.48$) 
and with $\rho(u,z)\approx 1.44$
we arrive at the small shift
\bes
  r_0 \Delta \Mbeauty &=& - 0.055(17) \,.  \label{e:shift1}
\ees 
A similar error is be taken into account
due to the 2\% uncertainty in the relation $L_2=1.436r_0$ \cite{pot:intermed}. 
In the same way it leads to a {\em statistical error} of
\bes
  r_0 \Delta \Mbeauty &=& 0.016\,.	 \label{e:shift2} 
\ees
The two contributions \eq{e:shift1}, \eq{e:shift2} are combined to
\bes
    r_0 \Delta \Mbeauty &=& - 0.055(23) \,,\label{e:shift} 
\ees
which we have taken into account in \sect{s:res3}. Because of the smallnes
of these effects, they can be neglected in the $\minv$-corrections.

In the case of our alternative strategy, the shift depdends on the value of
$\theta_0$. We find
\bes
&\theta_0=0   &\qquad   r_0 \Delta \Mbeauty  = -0.042(20) \,,\\
&\theta_0=1/2 &\qquad   r_0 \Delta \Mbeauty  =  0.009(11) \,,\\
&\theta_0=1   &\qquad   r_0 \Delta \Mbeauty  =  0.150(45) \,.
\ees


\end{appendix}

\bibliographystyle{h-elsevier}   
\bibliography{refs}           

\end{document}